\documentclass[fleqn,usenatbib]{mnras}

\usepackage{newtxtext,newtxmath}

\usepackage[T1]{fontenc}

\DeclareRobustCommand{\VAN}[3]{#2}
\let\VANthebibliography\thebibliography
\def\thebibliography{\DeclareRobustCommand{\VAN}[3]{##3}\VANthebibliography}

\usepackage{graphicx}	
\usepackage{amsmath}	
\usepackage[utf8]{inputenc}

\usepackage{csvsimple}
\usepackage{booktabs}
\usepackage{longtable}
\usepackage{pgf}
\usepackage{ifthen}

\usepackage{caption} 
\usepackage{orcidlink}

\usepackage{placeins}
\pgfkeys{/pgf/number format/.cd, fixed, fixed zerofill, precision=3}
\newcommand{\fmtnum}[1]{\ifthenelse{\equal{#1}{}}{---}{\pgfmathprintnumber{#1}}}

\title[Single-Transit Parameter Estimation]{Single-Transit Parameter Estimation with Calibrated Uncertainties}

\author[Helem Salinas et al.]{Helem Salinas$^{1}$\orcidlink{0009-0009-8795-4563}\thanks{E-mail: yhsalinas@uc.cl}, 
Rafael Brahm \orcidlink{0000-0002-9158-7315}$^{1,2,3}$, 
Greg Olmschenk$^{4,5}$\orcidlink{0000-0001-8472-2219},
Melinda Soares-Furtado$^{7,8,9}$\orcidlink{0000-0001-7493-7419},
\newauthor Karim Pichara$^{2,6}$,
Joshua E. Schlieder $^{4}$\orcidlink{0000-0001-5347-7062}
\\
$^{1}$Facultad de Ingeniería y Ciencias, Universidad Adolfo Ibáñez, Av. Diagonal Las Torres 2640, Peñalolén, 7941169, Santiago, Chile\\
$^{2}$Millennium Institute for Astrophysics, Av. Vicuna Mackenna 4860, 7820436 Macul, Santiago, Chile\\
$^{3}$Data Observatory Foundation, Eliodoro Yañez 2990, Providencia, 5110277, Santiago, Chile\\
$^{4}$NASA Goddard Space Flight Center, Greenbelt, MD 20771, USA\\
$^{5}$Department of Astronomy, University of Maryland, College Park, MD 20742, USA\\
$^{6}$Departamento de Ciencias de la Computación, Facultad de Ingeniería, Pontificia Universidad Católica de Chile, Santiago 7820436, Chile\\
$^{7}$Department of Astronomy, University of Wisconsin--Madison, 475 N.~Charter St., Madison, WI 53706, USA\\
$^{8}$Department of Physics, University of Wisconsin--Madison, 1150 University Ave, Madison, WI 53706, USA\\
$^{9}$Wisconsin Center for Origins Research, University of Wisconsin--Madison, 475 N Charter St., Madison, WI 53706, USA\\
}

\date{Accepted XXX. Received YYY; in original form ZZZ}

\pubyear{\the\year{}}

\begin{document}
\label{firstpage}
\pagerange{\pageref{firstpage}--\pageref{lastpage}}

\maketitle

\begin{abstract}
The growing volume of photometric data from current and upcoming exoplanet surveys demands automated analysis pipelines to detect and characterize exoplanet transit signals. Among these, single-transit events present a challenge, since only one event is observed, and standard transit-fitting techniques for periodic signals do not directly apply because they rely on phase-folding. Beyond detection, characterizing these events is equally critical for follow-up, yet existing approaches are either sensitive to stellar variability and require prior detrending, or depend on stellar priors and per-target fitting. This work presents a convolutional neural network that predicts transit depth, duration, and mid-transit time, together with calibrated uncertainties, directly from a single transit signal. Our network is trained on synthetic light curves and evaluated against a trapezoidal least-squares baseline. On 120 confirmed exoplanets, both methods achieve comparable median errors, but the baseline fails to locate the transit in 10\% of cases, producing unreliable estimates that would propagate into follow-up prioritization. In contrast, our network fails in only 1\% of cases, yields lower mean absolute errors, and does not require detrending or stellar priors. Detrending would distort or remove the transit along with high stellar variability, precisely where the baseline fails and our network succeeds. Finally, we identify 169 new transit candidates in TESS Sectors 27–60 and apply our approach to obtain their physical parameters, creating a catalog that includes single-transit, multi-transit and multi-planet systems. This approach offers a scalable path toward automated characterization of single transit events, a population that will grow with upcoming surveys.


\end{abstract}




\section{Introduction}
\label{sec:introduction}
The volume of photometric data from space missions such as Kepler \citep{borucki2010kepler} and the Transiting Exoplanet Survey Satellite \citep[TESS;][]{ricker2015transiting} has grown to a scale that demands automated end-to-end analysis pipelines. This applies to the detection of transiting candidates and to the characterization of their physical parameters. The standard approach to identifying and characterizing transiting candidates involves removing long-term stellar variability through a prior detrending step and searching for periodic signals by phase-folding the light curve (LC) to different trial periods. This is commonly implemented via the Box Least Squares algorithm \citep[BLS;][]{kovacs2002box}. This yields threshold crossing events (TCEs) that are then vetted to reject false positives such as eclipsing binaries (EBs) and instrumental artifacts \citep{tenenbaum2012detection}. Although BLS provides initial estimates of the period, depth, and duration as part of the search process, these are approximate values tied to the phase-folding grid and do not apply when only a single transit is observed in the LC. Moreover, since BLS requires a prior detrending step, this can distort or remove transit signals and alter the recovered transit parameters, which could increase the errors in the characterization stage \citep{kunimoto2022tess, eschen2024nine}. A single transiter, also referred to as a monotransit, occurs when the orbital period exceeds the observation baseline, such that only one transit event appears in the available LC, whether observed in a single sector or across several. Such events are intrinsically rare in photometric surveys, as long-period planets have a lower geometric transit probability and are less likely to transit within any given observation baseline \citep{osborn2016single}. Nevertheless, a growing number of single transits candidates have been identified in both Kepler and TESS data \citep{foreman2016population, gill2020ngts, hawthorn2024tess, magliano2024plato, salinas2025}, demonstrating that they represent a detectable and scientifically valuable population. \par

The discovery of single transiters is particularly valuable for characterizing long-period giant planets, which are still a poorly studied population of transiting systems, but crucial for understanding formation and migration processes. This is mainly because the transit probability $P_{\rm tr} \approx R_\star/a$ decreases with orbital separation $a$, making long-period planets geometrically less likely to transit and observationally harder to detect within the limited observing duration of current surveys \citep{winn2010exoplanet}. As a result, the vast majority of characterized transiting giants have orbital periods shorter than 10 d \citep{hartman2019hats}, a population dominated by hot Jupiters whose existence has long challenged standard theories of giant-planet formation \citep{pollack1996formation}. In contrast, long-period giants are less affected by intense stellar irradiation. Moreover, their orbital properties, in particular the eccentricity and obliquity distribution, help discriminate between competing migration scenarios \citep{dawson2018origins}. In this context, single transiters offer a way to access this population, and dedicated follow-up campaigns have already confirmed several tens of long-period transiting giants first identified as such events in TESS, with orbital periods ranging from tens to hundreds of days \citep{ulmer2022two, dalba2022tess, incha2023kepler, brahm2023three, mann2023giant, essack2025giant, kendall202643, rojas2026, bieryla2026toi}. Since each candidate demands its own extensive spectroscopic and photometric follow-up, fast and reliable transit-parameter estimates at the detection stage are essential to prioritize targets and plan observations efficiently. \par

For these events, physical parameters are typically estimated by fitting a parametric transit model \citep{mandel2002analytic} directly to the single transit event via Bayesian inference \citep[e.g.][]{osborn2016single, osborn2022monotools, magliano2024plato}, returning posterior distributions over depth, duration, and mid-transit time. Because single transits lack the signal-to-noise ratio (SNR) calculated by phase-folding multiple transits, the parameter estimation is particularly sensitive to local photometric noise and stellar variability, which can distort the transit shape \citep{foreman2016population}. To separate the transit signal from stellar variability, approaches such as \textsc{Allesfitter} \citep{gunther2021allesfitter} and \textsc{MonoTools} \citep{osborn2022monotools} model the transit and stellar variability jointly using Gaussian Processes (GPs), incorporating priors on stellar properties from the TESS Input Catalog \citep{stassun2019revised}, limb-darkening, and impact parameter. This allows the fit to distinguish a true transit from features induced by stellar variability, that would otherwise mimic or mask the true transit signal. However, these priors are not always reliable, as stellar parameters can carry systematic biases for faint or under-characterized stars \citep{stassun2019revised}. In addition, simultaneous modeling of the transit and the GP adds dimensionality and computational cost to the fit, further limiting the scalability of per-target inference. \par

A faster alternative for parameter estimation is local box-fitting without phase-folding, which adapts the core mechanism of the BLS algorithm \citep{kovacs2002box} to a single transit window. Building on this concept, newer approaches replace the box profile with trapezoidal least-squares (Trapezoidal-LS) or analytic transit models \citep{mandel2002analytic, li2019kepler} to accurately capture the ingress and egress phases \citep{kipping2023snr}. Fitting these models directly to the local transit window provides rapid parameter estimation without relying on priors or orbital periodicity. In practice, because it fits only the local transit window without modeling the out-of-transit flux, this local fitting approach is prone to absorbing stellar variability into the transit parameters or requires prior detrending, which risks removing the transit along with the variability. A related approach, Transit Least Squares (TLS; \citealp{hippke2019optimized}), incorporates limb darkening in the fitted profile, but is also typically applied after prior detrending, requires stellar priors as input, and is designed for periodic transits rather than single transit events. \par

Automated pipelines, including approaches that use deep learning \citep[DL;][]{lecun2015deep}, often rely on phase-folding, which requires prior knowledge of the orbital period and inherently excludes single transit signals from both detection and characterization \citep[e.g.,][]{osborn2020rapid, rao2021nigraha, tey2023identifying, salinas2023}. Notable exceptions are works that process raw, unfolded LCs directly: \citet{olmschenk2021identifying} developed a convolutional neural network pipeline \citep[CNN;][]{krizhevsky2012imagenet} that processes complete TESS Full-Frame Image (FFI) LCs, and \citet{salinas2025} introduced a transformer architecture that similarly operates on raw LCs and identified both multi-transit and single transit signals. In both cases, the network is trained on the entire LC, which allows it to learn not only from the transit itself but also from its context, without requiring prior removal of stellar variability. This reflects a more general property of a neural network (NN), which can learn to use sources of noise as informative context rather than removing them explicitly \citep{lecun2015deep}. Motivated by this, we characterize individual single transit events as the next stage of the pipeline, without requiring prior detrending or stellar priors. In this way, our approach treats instrumental artifacts and stellar variability as background variation. \par

In this work, we propose a CNN that predicts transit depth, duration, and mid-transit time, together with calibrated uncertainties, directly from a single transit signal. This NN is trained on synthetic LCs that include stellar variability, correlated noise, and realistic photometric noise, and does not require prior detrending or phase-folding. We evaluate performance using parameter prediction errors and epoch failure rates, which occur when a model fails to locate the transit center by a large margin, fitting out-of-transit stellar variability instead. Once trained, parameter inference has a constant computational cost per candidate, requiring only a single forward pass over a fixed input window. In contrast, grid-based Trapezoidal-LS scales directly with parameter grid resolution, while analytic transit fitting requires iterative optimization, where convergence speed varies per candidate. To assess its performance, we evaluate the CNN on synthetic test data through a series of sensitivity tests. Furthermore, we compared it with a Trapezoidal-LS baseline on 120 confirmed TESS exoplanets. We adopt Trapezoidal-LS as a baseline because it requires no limb-darkening dependencies or stellar parameters, thereby avoiding the parameter degeneracies and added computational cost of analytic fits on single transits. As a result, we find that both methods achieve comparable median errors, but the CNN yields lower mean absolute errors for depth and duration, particularly under high stellar variability, and achieves an epoch failure rate of about 1\%, compared with around 10\% for Trapezoidal-LS. This is because the CNN learns to distinguish the transit from stellar variability, whereas the Trapezoidal-LS fit tends to absorb variability features into the transit parameters. \par

This article is organized as follows: Section~\ref{sec:data_simulation} describes the synthetic transit simulation used to train and evaluate the CNN. Section~\ref{sec:methodology} presents the network architecture, uncertainty estimation, and experimental setup. In Section~\ref{sec:results}, we report the performance on synthetic data, including sensitivity tests and comparison with the Trapezoidal-LS baseline. Section~\ref{sec:application_real_tesslc} describes the 
application to real TESS LCs, including validation on confirmed planets. Then, Section~\ref{sec:new_candidates} describes the application of our network to new transit candidates from TESS Sectors 27--60. Section~\ref{sec:discussion} presents the discussion and illustrative candidates. Finally, we summarize our conclusions in Section~\ref{sec:conclusions}.

\section{Data: Synthetic Transit Simulation}
\label{sec:data_simulation}
In this section, we describe the process used to generate synthetic LCs for training and evaluating the CNN. Unlike real observations, which would require extracting each transit and estimating these quantities before use, limiting the diversity and difficulty of the training set, synthetic LCs allow us to vary stellar variability, noise, and transit properties independently. This includes challenging low-SNR and high-variability cases that help the network generalize and evaluate its performance across different scenarios. In our simulations, each LC combines a physical transit signal with a slowly varying photometric baseline and four noise components:

\begin{equation}
F_{\rm obs}(t) = F_{\rm transit}(t) \cdot B(t) + \epsilon_{\rm w}(t) + \epsilon_{\rm r}(t) + \epsilon_{\rm GP}(t) + \epsilon_{\rm out}(t)
\label{eq:f_t}
\end{equation}
where $F_{\rm transit}(t)$ is the transit signal, $B(t)$ is a slowly varying photometric baseline, and $\epsilon_{\rm w}$, $\epsilon_{\rm r}$, $\epsilon_{\rm GP}$ and $\epsilon_{\rm out}(t)$ are the white-noise, red-noise, stellar variability, and outlier components, respectively. \par

Time $t$ is measured from an approximate transit time $\hat{t}_0$, taken as the time reference, with $t = 0$ at $\hat{t}_0$. The true mid-transit time $t_0$ is offset from this reference by a random epoch offset $\Delta t_0 \equiv t_0 - \hat{t}_0 \sim \mathcal{U}(-40\,\mathrm{h},\,+40\,\mathrm{h})$, capturing the timing error introduced by the detection pipeline. This range ensures that the transit always falls within the $\sim$100 h input window while retaining sufficient out-of-transit flux on both sides. In this way, our network learns to recover $\Delta t_0$, along with the depth and duration. \par

To model the transit function, $F_{\rm transit}(t)$, we utilize the \texttt{batman} package \citep{kreidberg2015batman}. Since our goal is to predict the observed transit depth, duration and mid-transit time rather than orbital elements, we fix the orbital eccentricity to $e = 0$. Following \cite{seager2003unique}, the total transit duration, $t_{\rm T}$ (first to fourth contact), is given by:

\begin{equation}
t_{\rm T} = \frac{P}{\pi}\, \arcsin\!\left[\frac{\sqrt{(1+r_{\rm p})^2 - b^2}}{a_{\rm rs}\,\sin i}\right],
\label{eq:t14}
\end{equation}

where $P$ is the orbital period, $b$ is the impact parameter, $a_{\rm rs} \equiv a/R_\star$ is the scaled semi-major axis, and $i$ is the orbital inclination. The planet-to-star radius ratio $r_{\rm p} \equiv R_{\rm p}/R_\star$ is set by a target geometric depth $\delta_{\rm target} = r_{\rm p}^{2}$, sampled log-uniformly in ppm. The observed depth, defined as $\delta_{\rm obs} = 1 - \min[F_{\rm transit}(t)]$, is the quantity directly measurable from the LC and used as the depth target during training. \par

To ensure the physical viability of each simulated system, $b$ and $t_{\rm T}$ are sampled jointly. We sample $b \sim \mathcal{U}(0, b_{\max})$, with $b_{\max}=0.90$, excluding the most grazing geometries, where the transit profile becomes increasingly V-shaped and more characteristic of EBs than of planetary transits, and compute the inclination from:

\begin{equation}
i = \arccos\!\left(\frac{b}{a_{\rm rs}}\right),
\label{eq:inc}
\end{equation}

Then, we evaluate $t_{\rm T}$ from Eq.~\ref{eq:t14}. The resulting value is accepted only if $t_{\rm T}$ falls within the duration limits described in Table~\ref{tab:duration_parameters}, and is otherwise resampled.

For quadratic limb darkening, we adopt coefficients $(u_1, u_2)$ obtained using the triangular scheme, sampling $q_1, q_2 \sim \mathcal{U}(0,1)$ and transforming to $u_1 = 2\sqrt{q_1}\,q_2$ and $u_2 = \sqrt{q_1}\,(1-2 q_2)$ \citep{kipping2013efficient}. 

Table~\ref{tab:duration_parameters} summarizes the parameter ranges adopted for the transit model of the simulated dataset. The wide range parameters ($\delta_{\rm target}$, $a_{\rm rs}$) are sampled log-uniformly to balance coverage across scales, and the remaining parameters uniformly.

\begin{table*}
\centering
\caption{Summary of parameters.}
\label{tab:duration_parameters}
\begin{tabular}{lcccll}
\hline \hline
\textbf{Parameter} & \textbf{Minimum} & \textbf{Maximum} & \textbf{Unit} & \textbf{Prior} & \textbf{References}\\ \hline
&&&\\
Orbital period ($P$) & 12.0 & 200.0 & d & uniform &\citet{seager2003unique}\\
Scaled semi-major axis ($a_{\rm rs}$) & 8.0   & 60.0 & --- & log-uniform & \citet{seager2003unique, mullally2015planetary} \\
Impact parameter ($b$) & 0.0   & 0.90 & --- & uniform & \citet{mandel2002analytic} \\
Transit depth ($\delta_{\rm target}$) & 300 & 15,000 & ppm & log-uniform &\citet{fressin2013false, wilson2023transiting} \\
Transit duration ($t_{\rm T}$) & 1.5   & 15.0 & h   & derived & \citet{seager2003unique}\\
Orbital eccentricity ($e$) & \multicolumn{2}{c}{0.0 (fixed)} & --- & --- & \citet{kipping2008transiting, kreidberg2015batman} \\
Epoch offset ($\Delta t_0$) & $-40$ & $+40$ & h & uniform&---  \\
\hline
&&&\\
Linear slope ($s$)  & $-200$ & $+200$  & ppm\,d$^{-1}$  & uniform & --- \\
Quadratic curvature ($q$)& $-20$& $+20$ & ppm\,d$^{-2}$ & uniform & ---\\
\hline
&&&\\
White-noise std.\ ($\sigma_w$) & 50   & 1500 & ppm & log-uniform & \citet{ricker2015transiting} \\
Red-noise std.\ ($\sigma_r$) & 10   & 200 & ppm & log-uniform &\citet{caceres2019autoregressive} \\
Autoregressive noise coefficient ($\phi$) & \multicolumn{2}{c}{0.85 (fixed)} & --- & --- & \citet{caceres2019autoregressive} \\
\hline
&&&\\
Amplitude ($\sigma_\star$) & 50   & 1500  & ppm & log-uniform &\citet{haywood2014planets, mcquillan2014rotation}\\
Rotation period ($P_{\rm rot}$) & 2.0  & 30.0& d & uniform &\citet{haywood2014planets, mcquillan2014rotation}\\
Active-region evolution timescale ($\ell$) & 0.5  & 10.0   & d   & uniform & \citet{haywood2014planets, foreman2017fast}\\
Periodicity factor ($\alpha$) & 0.3& 1.0& --- & uniform & \citet{angus2018inferring}  \\

\hline \hline
Outlier amplitude ($\sigma_{\rm out}$) & \multicolumn{2}{c}{5000 (fixed)} & ppm & --- & --- \\
Outlier probability ($p_{\rm out}$) & \multicolumn{2}{c}{$5\times 10^{-4}$ (fixed)} & --- & --- & ---\\
\hline \hline
\end{tabular}
\end{table*}

The remaining components of Eq.~\ref{eq:f_t} describe the photometric baseline and the noise sources that we add to the transit signal. The baseline $B(t)$ accounts for slow photometric drifts and is modeled as a low-order polynomial in time:

\begin{equation}
B(t) = 1 + \bigl(s\,t + q\,t^{2}\bigr)\times 10^{-6},
\label{eq:baseline}
\end{equation}
where $t$ is measured from the center of the observation window, $s$ is the linear slope (in ppm per day) and q is the quadratic curvature (in ppm per day$^2$). The order $N = 2$ is the minimum sufficient to capture both drift and curvature on multi-day timescales, consistent with \citet{wong2020systematic, wong2021visible}. For each LC, $s$ and $q$ are sampled uniformly from symmetric intervals around zero, with limits given in Table~\ref{tab:duration_parameters}.

Of the four noise components, first, the white-noise component $\epsilon_{\rm w}(t)$ represents the uncorrelated photometric scatter set by photon counting and detector read noise. We model it as independent Gaussian samples at each cadence:
\begin{equation}
    \epsilon_{\rm w}(t) \sim \mathcal{N}(0,\,\sigma_{\rm w}^{2}),
\end{equation}
with the per-point standard deviation $\sigma_w$ sampled log-uniformly in [50, 1500] ppm per 30-minute cadence, matching the sampling of TESS FFIs and covering the noise range expected for typical exoplanet search targets \citep{ricker2015transiting}. Second, $\epsilon_{\rm r}(t)$ represents the time-correlated short-term red noise, which is treated as a first-order autoregressive process \citet{caceres2019autoregressive}, to inject realistic correlated noise into our training data. It is defined as:

\begin{equation}
\epsilon_{\rm r}[i] = \phi\,\epsilon_{\rm r}[i-1] + \eta_i, \qquad \eta_i \sim \mathcal{N}(0, \sigma_\eta^2),
\label{eq:ar1}
\end{equation}
where $\phi$ is the autoregressive correlation coefficient and $\eta_i$ is an uncorrelated Gaussian innovation at time step $i$.

Third, $\epsilon_{\rm GP}(t)$ captures long-term correlated stellar variability, such as stellar rotation or spot modulation. We model it as a GP with the quasi-periodic covariance kernel \citep{haywood2014planets}, a standard model for rotationally modulated stellar variability in LCs \citep{aigrain2015precise, angus2018inferring}:

\begin{equation}
k(\tau) = \sigma_\star^{2} \exp\!\left[-\frac{\tau^{2}}{2\ell^{2}} - \frac{\sin^{2}(\pi\tau/P_{\rm rot})}{2\alpha^{2}}\right],
\label{eq:qp_kernel}
\end{equation}
where $\tau \equiv t_i - t_j$ is the time lag between two points, $\sigma_\star$ is the variability amplitude, $\ell$ is the active-region evolution timescale, $P_{\rm rot}$ is the stellar rotation period, and $\alpha$ controls the distinctness of the periodic structure \citep{haywood2014planets, angus2018inferring}. The parameter ranges adopted for $P_{\rm rot}$, $\ell$, and $\alpha$ are listed in Table~\ref{tab:duration_parameters}. The rotation period range covers the most populated region of the distribution observed for FGK main-sequence stars \citep{mcquillan2014rotation}, and the evolution timescale range spans from rapidly evolving active regions to regions coherent over the $\sim$100 h input window \citep{haywood2014planets}.\par

 

Finally, we add photometric anomalies caused by cosmic-ray hits, momentum dumps, and residual detector artifacts \citep{jenkins2010overview, ricker2015transiting}. We model these instrumental systematic effects as a Bernoulli-GP, following the noise modeling strategy of \citet{wunderlich2023data} where random Gaussian outliers are defined as:

\begin{equation}
    \epsilon_{\rm out} \sim \mathcal{N}(0, \sigma_{\rm out}^2)
\end{equation}
and are injected at each cadence with a fixed Bernoulli probability $p_{\rm out}$. The values of $p_{\rm out}$ and $\sigma_{\rm out}$ are listed in Table~\ref{tab:duration_parameters}. This exposes the NN to rare high-amplitude points, which helps it to handle isolated anomalies. Figure~\ref{fig:dataset_sim} shows some examples of synthetic LCs generated with this procedure.

\section{Method}
\label{sec:methodology}
In this section, we describe our methodology, which is designed to estimate transit parameters: depth, duration, and mid-transit time, together with calibrated uncertainties, directly from a single transit signal. Our work is motivated by the need for fast, scalable characterization of single transit events that cannot be handled by phase-folding techniques. Our methodology consists of four main steps, which we describe in the following subsections. First, in Section~\ref{sec:data_preprocessing}, we describe the data preprocessing. Then, we present the NN architecture in Section~\ref{sec:neural_network_architecture}, the uncertainty estimation in Section~\ref{sec:uncertainty_estimation}. Finally, the experimental setup in Section~\ref{sec:experimental_setup}.

\subsection{Data preprocessing}
\label{sec:data_preprocessing}
The LCs are preprocessed in a consistent format across both training and inference. Each input corresponds to a single window of 99.5 hours built around an approximate transit time $\hat{t}_0$, sampled on a fixed grid of 200 points at a 30-minute cadence. During training, we used synthetic LCs, and $\hat{t}_0$ is provided by our transit simulation. For inference, we use real TESS LCs and $\hat{t}_0$ is provided by a prior detection pipeline. For real LCs, we binned them to a 30-minute cadence, and we linearly interpolated onto this grid, which also fills missing cadences and data gaps within the window. Since interpolation across a gap that overlaps the transit can bias the predicted parameters, we visually inspect such cases. \par

\subsection{Neural network architecture}
\label{sec:neural_network_architecture}
\begin{figure*}
  \centering
  \includegraphics[width=0.7\linewidth]{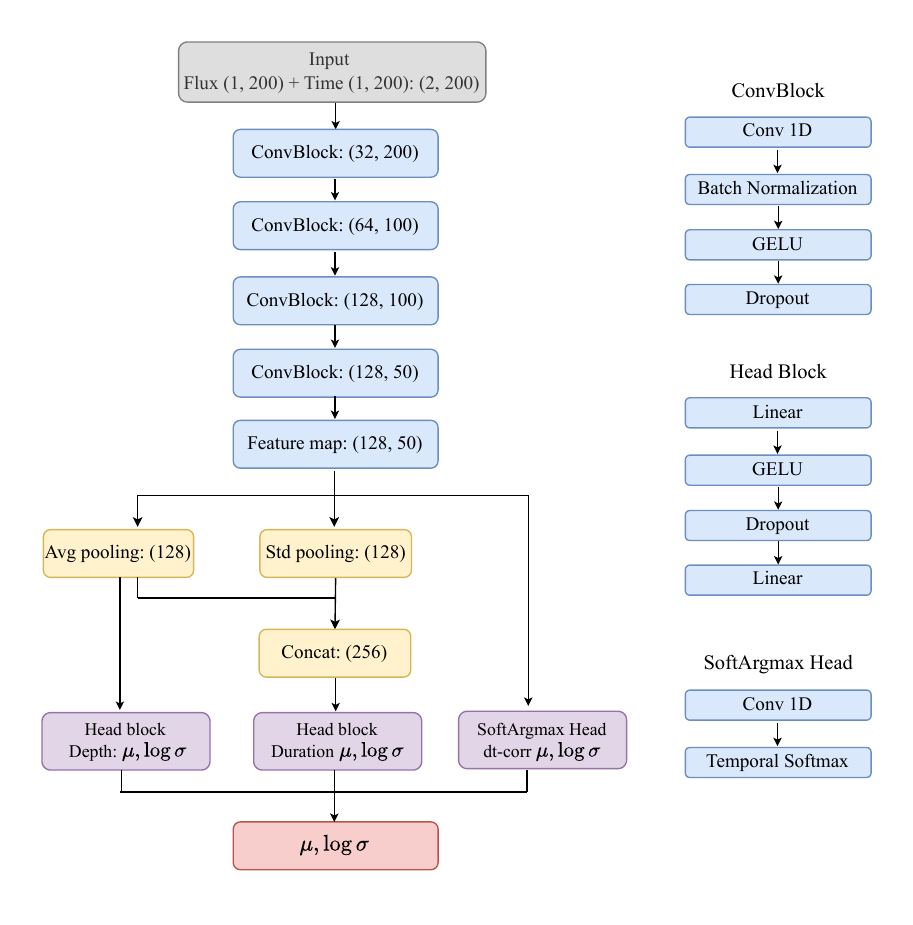}
\caption{An overview of the architecture of the CNN used in this work for transit parameter estimation. The network passes input LC through a four-layer 1D convolutional network. For depth, we use average pooling, while for duration, we concatenate average and standard deviation pooling. For mid-transit time, the feature map passes directly to a SoftArgmax head. For each parameter, the network predicts $\mu$ and the log-uncertainty $\log{\sigma}$.}  
    \label{fig:scheme_cnn}
\end{figure*}

Our proposed architecture, illustrated in Figure~\ref{fig:scheme_cnn}, is a CNN designed to operate directly on the single transit signal. It consists of four 1D convolutional blocks that progressively extract features at increasing temporal scales, followed by parameter-specific pooling heads that produce the final predictions.\par

We adopt a 1D convolutional architecture because transit signals contain local features that convolutional filters can detect across multiple temporal scales through hierarchical pooling \citep{lecun2015deep}. Since exact transit times are frequently unreported or imprecise in detection pipelines, resulting in an epoch offset $\Delta t_0$, the transit can appear at any position within a given window. Consequently, the same convolutional filters are applied across the full sequence of the signal regardless of the transit location.\par

The NN takes as input an LC window extracted around the approximate transit epoch, consisting of normalized flux and time, from which it learns to characterize the transit signal and recover depth, duration, and mid-transit time. The input passes through four 1D convolutional blocks with channel depths of 32, 64, 128, and 128. Each block comprises a convolution to extract local features from the LC. The first block preserves the temporal resolution, while the remaining three progressively reduce the number of time steps from 200 to 100 and finally to 50. In particular, the third and fourth blocks use dilated convolutions to capture features over longer timescales. Together, these blocks compress the temporal dimension and increase the receptive field of the NN, with early layers detecting fine-grained local features and deeper layers integrating information over progressively longer timescales. Then, we apply a batch normalization to normalize the output of convolutional layers. This normalization ensures that the distribution of activations remains consistent across layers, allowing the NN to learn more efficiently \citep{ioffe2015batch}, thus improving training convergence. The next layer consists of a Gaussian Error Linear Unit \citep[GELU;][]{hendrycks2016gaussian} activation function, which introduces non-linearity and enables the NN to learn complex mappings between the extracted features. Finally, the output of the GELU activation is passed to a dropout layer which randomly deactivates a fraction of neurons during training \citep{srivastava2014dropout}. This forces the NN to learn more generalized representations of the transit features, making it more robust to noise and variability in the input LCs, and reduces overfitting. \par                                                                                                

The resulting feature map is then summarized through parameter-specific pooling operations. The first applies average pooling to produce a global representation used to predict transit depth, which depends on the overall structure of the transit signal. The second concatenates this layer with the output of standard-deviation pooling to predict the transit duration \citep{snyder2018x}. This combination provides information not only on the mean level of the feature map, but also on how it varies across the transit signal, allowing the network to capture the transit duration independently of its depth. Unlike depth and duration, the mid-transit time depends on the position of the transit within the window, which is why we apply a soft-argmax head operating directly on the feature map, without pooling. This head preserves the positional information, allowing the correction to be obtained directly from where the transit appears on the feature map, yielding a more robust estimate. In this way, each parameter is predicted by a dedicated head. The resulting predictions are concatenated into a three-dimensional output vector, $\mu$, corresponding to depth, duration, and mid-transit time. \par

\subsection{Uncertainty estimation}
\label{sec:uncertainty_estimation}
Our NN is trained to predict both the mean and uncertainty of each transit parameter simultaneously. Rather than assuming a fixed noise level, we adopt a heteroscedastic formulation in which the predicted uncertainty varies with the input. This reflects the fact that different LCs have different noise levels and SNRs. Specifically, for each parameter, the NN outputs a mean $\mu$ and a log-uncertainty $\log{\sigma}$, which are optimized using the loss function described in Section~\ref{sec:loss}.

\subsubsection{Heteroscedastic loss}
\label{sec:loss}
Each target parameter is modeled as a Gaussian random variable, with mean and standard deviation predicted directly by the NN from the input window LC \citep{nix1994estimating, kendall2017uncertainties}. For numerical stability and to guarantee positivity of the predicted standard deviation, the NN outputs $\log\sigma_{id}$ rather than $\sigma_{id}$ directly, with $\sigma_{id}$ recovered by exponentiation:

\begin{equation}
  \mathcal{L} = \frac{1}{N} \sum_{i=1}^{N} \sum_{d=1}^{3} \left[ \frac{(y_{id} - \mu_{id})^2}{2\sigma_{id}^2} + \log\sigma_{id} \right],  
\end{equation}
where $y_{id}$ is the true value, $\mu_{id}$ is the predicted mean and $\sigma_{id}$ is the predicted standard deviation for sample $i$ and output parameter $d$. The loss $\mathcal{L}$ is evaluated in a normalized target space where each output parameter (depth in ppm, duration, and mid-transit time in hours) is standardized to zero mean and unit variance using statistics computed on the training set, ensuring the joint loss is not dominated by the parameter with the largest numerical range. The first term penalizes inaccurate predictions weighted by the inverse variance, and the second term acts as a regularizer on the predicted uncertainty. After inference, the predicted means and uncertainties are converted back to physical units by inverting the standardization applied to the training targets. \par

\subsubsection{Uncertainty calibration}
\label{sec:calibration}
Although the heteroscedastic loss encourages calibrated uncertainties during training, the predicted standard deviations may still be systematically over- or under-estimated when evaluated on the validation set \citep{lakshminarayanan2017simple}. To correct this, we adopt a post-training calibration step similar to \citet{guo2017calibration}. This step introduces a learnable scale parameter for each output parameter $d$ optimized by minimizing the same negative log-likelihood on the validation set with the main network weights kept fixed, and is given by:

\begin{equation}
    \log\hat{\sigma}_{id} = \log\sigma_{id} + s_d,
\end{equation}
where $s_d$ is the learned scale. This correction adjusts the predicted uncertainties without modifying the point estimates, and is applied at inference time to all predictions. The quality of the calibration is evaluated using the coverage probability at the 68 and 95 per prediction levels (Cov68 and Cov95), which correspond to the $1\sigma$ and $2\sigma$ intervals of a Gaussian distribution \citep{kuleshov2018accurate}.

\begin{figure*}
  \centering
  \includegraphics[width=1\linewidth]{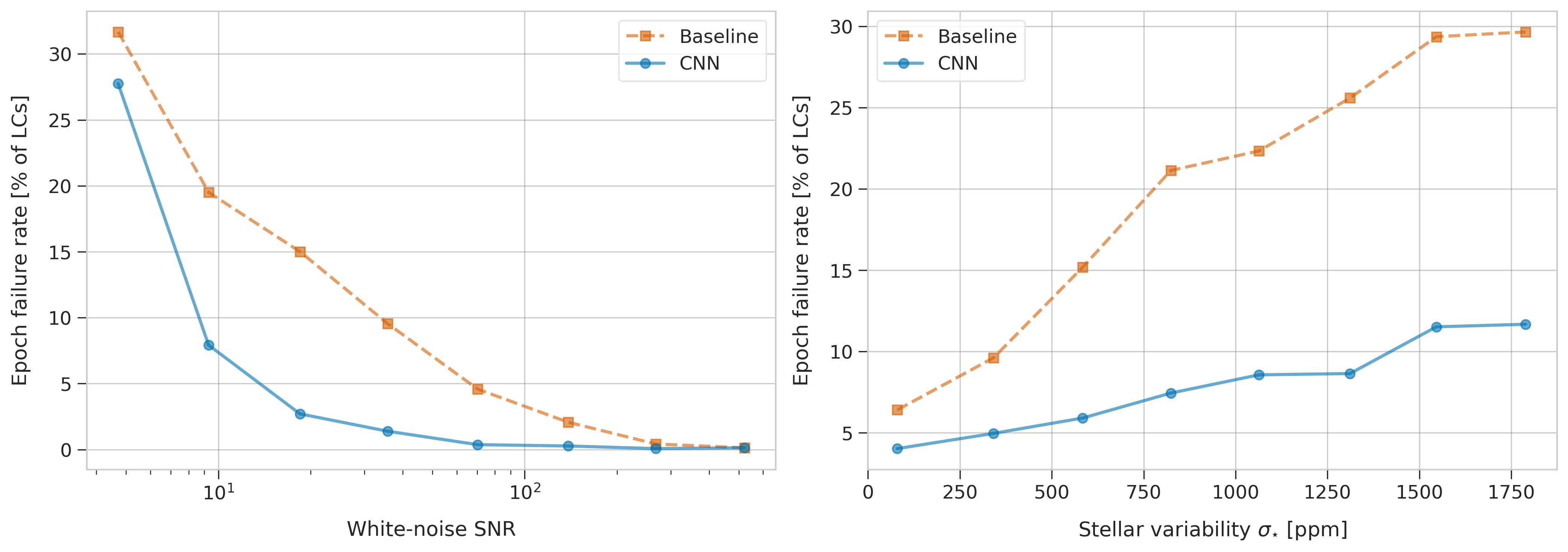}
\caption{Epoch failure rate on the simulated data as a function of white-noise SNR (left) and stellar variability amplitude $\sigma_\star$ (right) for the CNN and the Trapezoidal-LS baseline.}
    \label{fig:comparison_snr_stellar_combined}
\end{figure*}


\begin{table}
\centering
\caption{CNN performance on the evaluation test set of 45,000 synthetic LCs after uncertainty calibration. Cov68 and Cov95 are the coverage probabilities at the 68 and 95 per cent prediction intervals, respectively.}
\label{tab:test_metrics}
\begin{tabular}{lccc}
\hline  \hline
Parameter & MAE & Cov68 & Cov95 \\
\hline
Depth [ppm] & 170 & 0.718 & 0.947 \\
Duration [h]    & 0.88 & 0.703 & 0.955 \\
$\Delta t$ [h]  & 1.53 & 0.732 & 0.980 \\
\hline
\end{tabular}
\end{table}

\begin{figure}
  \centering
  \includegraphics[width=1\linewidth]{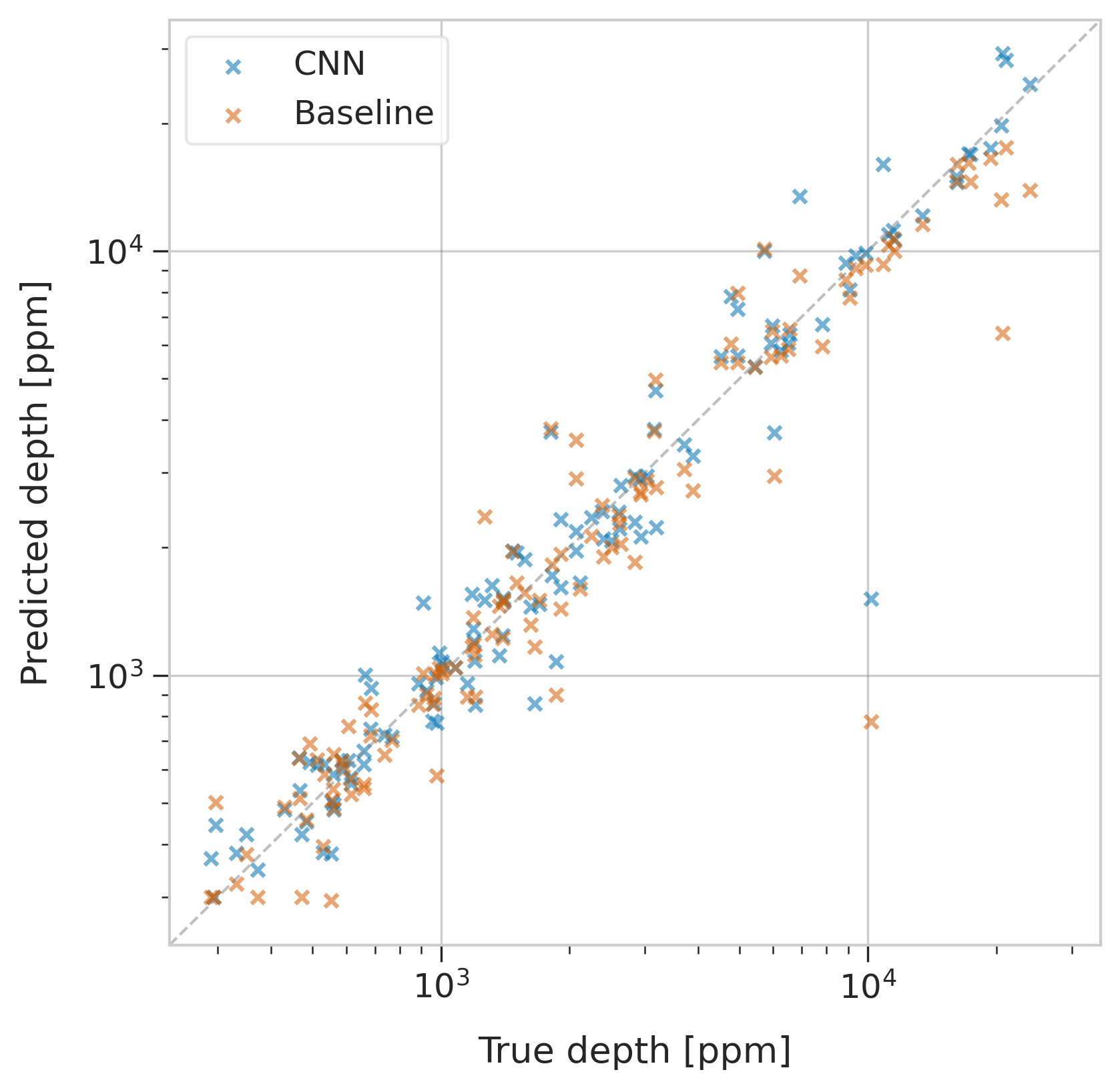}
\caption{Comparison between predicted and true transit depth for confirmed TESS planets, obtained with the CNN and the Trapezoidal-LS baseline. The dashed line indicates the one-to-one relation.} 
    \label{fig:depth_pred_vs_true}
\end{figure}
 
\begin{figure}
  \centering
  \includegraphics[width=0.9\linewidth]{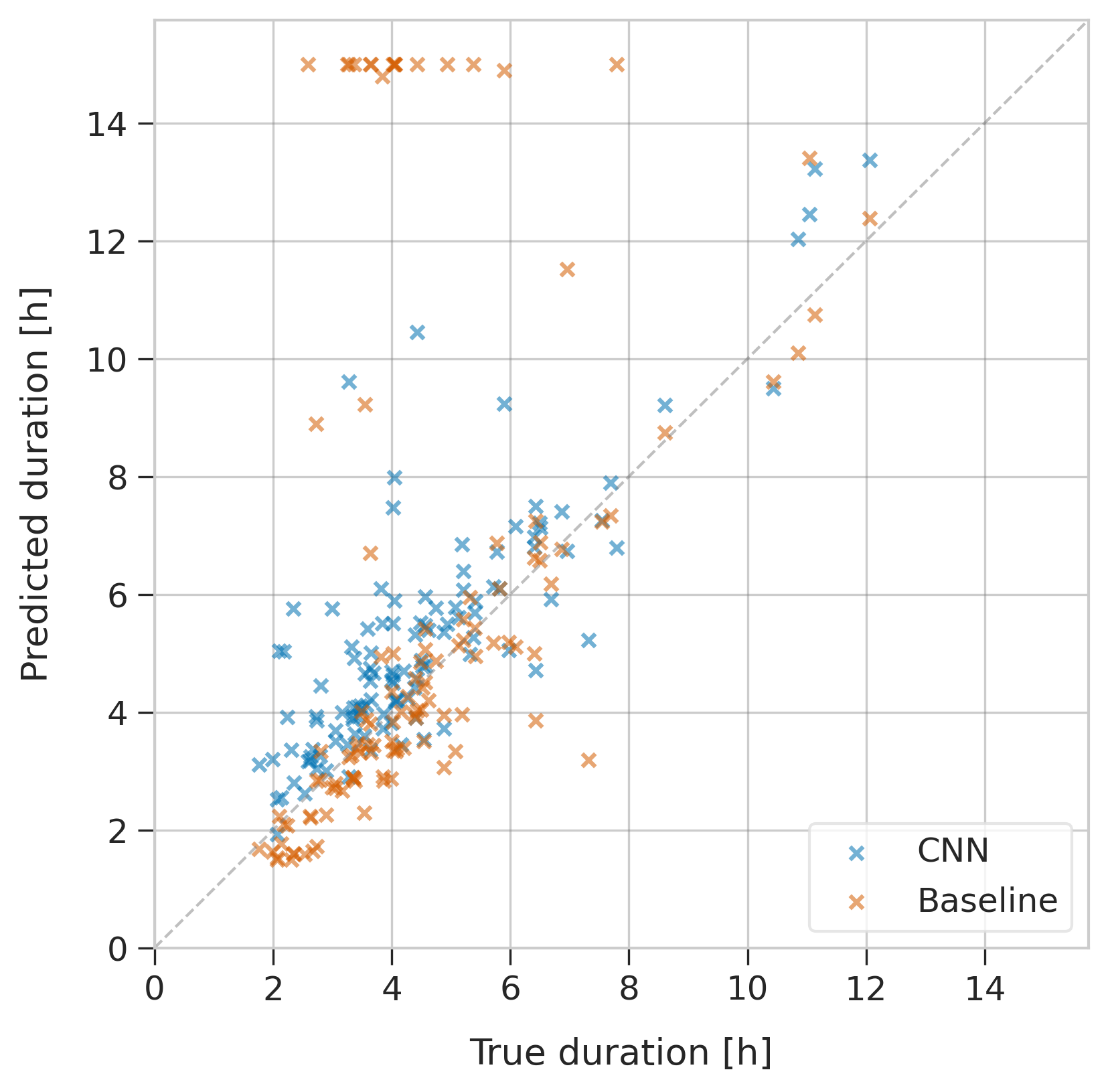}
\caption{Comparison between predicted and true transit duration for confirmed TESS planets, obtained with the CNN and the Trapezoidal-LS baseline. The dashed line indicates the one-to-one relation. Points saturating at the grid boundaries in the baseline correspond to catastrophic failures where the fit converges on stellar variability features rather than the true transit.}
  \label{fig:dur_pred_vs_true}
\end{figure}

\subsection{Experimental setup}
\label{sec:experimental_setup}
Our model, detailed in this work, was developed using \textsc{PyTorch}, an open-source framework. The code, together with the trained model, is publicly available on GitHub\footnote{\url{https://github.com/helemysm/singletransiters}}.

\subsubsection{Regularization and optimizers}
We use the AdamW optimizer, \citep{loshchilov2017decoupled}, which extends Adam \citep{kingma2014adam} with decoupled weight decay regularization, improving generalization particularly for models trained on large datasets. To initialize the training, we use a learning rate of $3 \times 10^{-4}$ and a weight decay of $10^{-4}$ \citep{bengio2012practical}. The learning rate is then reduced by factor of 0.5 when the validation loss shows no improvement for 15 consecutive epochs, with a minimum value of $10^{-6}$ to prevent the learning rate from becoming negligibly small \citep{lecun2015deep}. This allows the NN to continue optimizing its parameters as training progresses. During backpropagation process, we rescale the global gradient norm to a maximum value of 2.0 when exceeded. This prevents gradient updates from becoming excessively large and destabilizing the training process. Additionally, we employ regularization through the dropout layers described in Section~\ref{sec:neural_network_architecture}, which prevent the network from overfitting to the synthetic training distribution.

\subsubsection{Training details}
We generated 300,000 synthetic LCs using the process described in Section~\ref{sec:data_simulation} divided into 75 per cent for training, 10 per cent for validation, and 15 per cent for testing. We use the validation set both for learning rate scheduling and model selection. Our experiments were performed over 300 epochs with a batch size of 512, corresponding to approximately 440 gradient updates per epoch.

\section{Results}
\label{sec:results}

In this section, we present the performance of the CNN on synthetic test data and real TESS LCs. We first evaluate parameter recovery on our evaluation set, then characterize how performance varies with photometric noise, epoch offset, transit shape, and stellar variability through a series of sensitivity tests. Finally, we compare our NN against the Trapezoidal-LS baseline using confirmed TESS planets. \par

For all experiments and model evaluations, we adopt the mean absolute error (MAE) as our primary performance metric, as it provides a directly interpretable measure of typical prediction accuracy in physical units and is robust to outliers, in contrast to the mean squared error (MSE) which is dominated by large individual errors. We also assess uncertainty calibration through the coverage probability at Cov68 and Cov95, defined as the fraction of test samples for which the true parameter value falls within the predicted interval. For a perfectly calibrated Gaussian uncertainty, the expected coverage is 0.68 and 0.95 respectively. \par

\subsection{Parameter recovery}
We evaluated our NN on the evaluation set of 45,000 synthetic LCs. Table~\ref{tab:test_metrics} reports the MAE and coverage for each output parameter after uncertainty calibration. As a result, our NN recovers transit depth with MAE of 170 ppm, duration with MAE of 0.88 h, and mid-transit time with MAE of 1.53 h. The measured coverage at the 68 per cent level is 0.718, 0.703, and 0.732 for depth, duration, and mid-transit time respectively, close to the expected value of 0.68, indicating well-calibrated uncertainties on the synthetic test set. At the 95 per cent level, the coverage is 0.947, 0.955, and 0.980, slightly above the nominal expectation of 0.95, suggesting mildly conservative uncertainties at this level. \par

\subsection{Transit parameter sensitivity}
To characterize how the NN performance varies with observing conditions and transit properties, we evaluate the model on a series of controlled sensitivity sets. Each set varies one parameter at a time while holding all others fixed at training-realistic values, isolating the effect of each factor on parameter recovery. Each sensitivity set contains 2000 synthetic LCs.\par

\subsubsection{Sensitivity tests}
\label{sec:sensitivity_test}

We used five sensitivity tests: three photometric noise components (white noise $\sigma_w$, red noise $\sigma_r$, and quasi-periodic stellar variability $\sigma_\star$), the epoch offset $\Delta t_0$, and the impact parameter $b$. The ranges for each factor follow the distribution described in Table \ref{tab:duration_parameters}. For the white noise test, we evaluated $\sigma_w$ across discrete noise levels while setting $\sigma_r = 0$ and $\sigma_\star = 0$ to isolate the effect of white noise. For the remaining four tests, we sampled $\sigma_w$ log-uniformly in $[50, 3000]$ ppm per sample (median $\approx 325$ ppm). Thus, our NN is evaluated under realistic noise conditions. We summarize the performance at each level using the MAE of the three predicted parameters (depth, duration, and $\Delta t$). For $\Delta t$, we use the failure rate, i.e., the fraction of samples with $|\Delta t\text{ error}| > 5$ h. \par

We find that $\sigma_w$ produces the strongest degradation across all three parameters: as median SNR drops from $\sim$150 to $\sim$6, depth MAE grows gradually from $\sim$60 to $\sim$250 ppm up to $\sigma_w =$ 800 ppm. At higher noise levels, the error accelerates to $\sim$1130 ppm at $\sigma_w= $ 3000 ppm. This value corresponds to a median SNR $\sim$6, where the transit depth becomes comparable to the noise level. The $\Delta t$ failure rate follows a similar pattern, increasing from below 1\% to 16\%. For $\sigma_r$, we observe a similar effect, with the $\Delta t$ failure rate grows from 3.5\% to 13.1\% as $\sigma_r$ increases from 0 to 200 ppm, while depth MAE grows modestly from $\sim$206 to $\sim$268 ppm. \par 

At fixed SNR, $\sigma_\star$ is the most damaging single factor, the $\Delta t$ failure rate increases from 3.2\% ($\sigma_\star = 0$) to 15\% ($\sigma_\star = 1500$ ppm) while median SNR remains near 25. This performance indicates that quasi-periodic stellar modulations can shift the apparent transit center even when the transit itself remains detectable.

For the epoch offset test, we find that depth and duration MAE remain stable across all levels, while the $\Delta t$ failure rate only increases at offsets beyond $\Delta t_{\max} = 24$ h (from 3.7\% to 9.4\% at 40 h). The stability of depth and duration confirms that the SoftArgmax head absorbs the transit displacement without contaminating the other predictions. Finally, for the impact parameter $b$ test, depth MAE remains near $\sim$230 ppm across all levels. Duration MAE remains near $\sim$0.85 h for $b_{\max} \leq 0.85$ but jumps to 1.15 h at $b_{\max} = 0.95$, where V-shaped transits make the ingress and egress harder to resolve. The $\Delta t$ failure rate follows a similar pattern, increasing from $\sim$5\% to 9.7\% only at the most grazing geometries. We provide the full per-level results in Appendix \ref{ap:appendix}, Table~\ref{tab:sensitivity}.

\subsubsection{Baseline comparison}
\label{sec:baseline_comparison}
We compare our CNN against a Trapezoidal-LS baseline. This approach replaces the step-function box profile of the standard BLS algorithm \citep{kovacs2002box} with a symmetric trapezoid with linear ingress and egress, fitted directly on a single window to better approximate the transit profile. The trapezoid parameters are initialized from a grid search over the full window and then refined with a Nelder–Mead minimization \citep{nelder1965simplex}. We constrain the Trapezoidal-LS to the same depth, duration, and offset ranges used during CNN training (Table \ref{tab:duration_parameters}). Thus, both methods share the same parameter search ranges. We adopt this baseline as a fast classical estimator that, like our CNN at inference, requires no limb-darkening coefficients or stellar parameters, which may be unavailable or biased for faint or poorly characterized stars \citep{stassun2019revised}. The Trapezoidal-LS assumes a geometric profile with a uniform depth during transit, which means that it does not capture the limb-darkening curvature present in real transits. Although TLS \citep{hippke2019optimized} incorporates limb darkening, it is designed for periodic transits and requires stellar parameters (Section~\ref{sec:introduction}), which limits its application to single transit events. Our CNN, on the other hand, is trained on LCs generated using batman \citep{kreidberg2015batman} with quadratic limb darkening and thus implicitly incorporates this effect. \par

We evaluated both methods on 20,000 synthetic LCs with white noise, red noise, and stellar variability all varying simultaneously. To stress our NN beyond its training range, we extend $\sigma_w$ and $\sigma_\star$ up to 2000 ppm in this test set. Figure~\ref{fig:comparison_snr_stellar_combined} shows the $\Delta t$ failure rate as a function of white-noise SNR and stellar variability for both methods. The Trapezoidal-LS exhibits a higher failure rate than the CNN across the SNR range, with both methods converging to near zero only at the highest SNR. Across the full range of stellar variability amplitudes, the Trapezoidal-LS failure rate is roughly 2 to 3 times higher than that of the CNN. For duration, both methods degrade with decreasing SNR, whereas the depth MAE of the Trapezoidal-LS increases at high SNR. The advantage of our NN in both depth and duration grows when $\sigma_\star$ increases, suggesting that out-of-transit modulation degrades the Trapezoidal-LS fit (see Appendix~\ref{ap:appendix}, Figure~\ref{fig:mae_depth_dur_cnn_vs_ls}). \par

The cases where Trapezoidal-LS baseline fails are due to the fit interpreting stellar variability or red noise features as a transit signal, converging far from the actual transit location. When this happens, both the estimated depth and duration deviate significantly from the true values (Appendix~\ref{ap:appendix}, Figure~\ref{fig:prediction_sim_data}). This is one of the main reasons why traditional BLS pipelines require a prior detrending step, without it, stellar variability contaminates the phase-folded signal, degrading transit detection. Furthermore, detrending itself can distort or remove shallow transit signals, leading to missed detections or a high rate of false positives \citep{kunimoto2022tess, eschen2024nine}. In contrast, NNs learn to handle stellar variability during training, which not only avoids this limitation but also helps distinguish true transits from false positives \citep{olmschenk2021identifying, salinas2025}.

\begin{table}
\centering
\caption{Comparison of CNN and Trapezoidal-LS performance on depth and duration estimation across 120 confirmed planets with a $P$ $>10 $ days}
\label{tab:cnn_vs_ls}
\begin{tabular}{lcc}
\hline  \hline
\textbf{Metric} & \textbf{CNN} & \textbf{Trapezoidal-LS baseline} \\
\hline
Depth median error & 201 ppm & 196 ppm \\
Depth MAE & 717 ppm & 788 ppm \\
\hline
Duration median error & 0.65 h & 0.52 h \\
Duration MAE & 1.12 h & 1.92 h \\
\hline
$\Delta t$ median error & 0.31 h & 0.22 h \\
$\Delta t$ MAE & 0.87 h & 3.77 h \\
Failed fits (error $>$5 h)  & 2 / 120 (2\%) & 15 / 120 (13\%) \\
Failed fits (error $>$10 h)  & 1 / 120 (1\%) & 12 / 120 (10\%) \\
\hline
\end{tabular}
\end{table}

\section{Application to Real TESS Light Curves}
\label{sec:application_real_tesslc}
In this section, we describe the application of our NN to real TESS LCs. This analysis is divided into two parts: first, a validation on known planets from the TESS Objects of Interest (TOI) catalog, and second, the evaluation of new planetary candidates.

\begin{figure*}
  \centering
  \includegraphics[width=1.\linewidth]{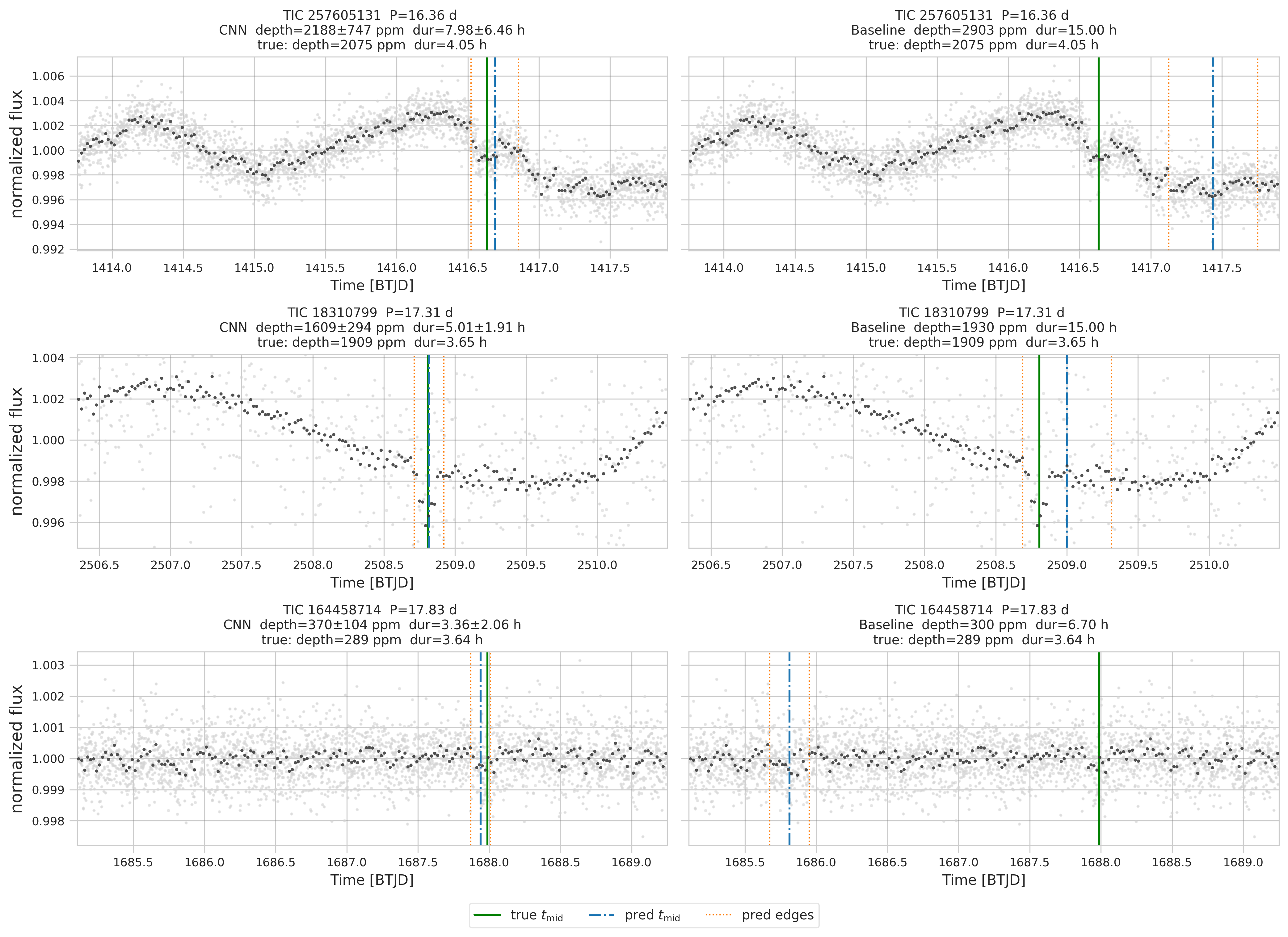}
\caption{Comparison of CNN and Trapezoidal-LS predictions for three confirmed TESS planets. Left column: CNN predictions with $1\sigma$ uncertainties. Right column: Trapezoidal-LS baseline predictions. In each panel, solid vertical lines mark the true mid-transit time, dot-dashed lines the predicted mid-transit time, and dotted lines the predicted transit edges. Predicted depth and duration, together with the true values, are annotated above each panel.}  
    \label{fig:result_confirmed_planet}
\end{figure*}

\subsection{Validation on known planets}
\label{sec:validation_know_planets}
We evaluated our NN on confirmed TESS exoplanets from the TOI catalog, selected as those with a TFOPWG disposition of Confirmed Planet (CP) or Known Planet (KP) and orbital period $P$ > 10 days. As reference values for the depth, duration, and mid-transit time, we adopt those listed in the TOI catalog. For each planet, we used SPOC LCs, and QLP LCs where SPOC data are unavailable. Then, we normalize the flux by its median and offset it by subtracting unity, effectively centering the out-of-transit flux at zero. To evaluate the robustness of the NN in recovering the correct transit center, we apply a random epoch jitter drawn uniformly from $[-$25, $+$25$]$ hours to the window center. In total, we evaluated 120 confirmed planets.\par

Consistent with our baseline comparison, we evaluated the Trapezoidal-LS baseline under the same grid constraints as the CNN. This restriction is necessary because an unconstrained Trapezoidal-LS grid can yield extreme, non-physical values for depth or duration in low signal-to-noise windows, which could inflate the overall MAE calculations. We only extended the upper depth limit of the Trapezoidal-LS grid to 25000 ppm to prevent it from saturating at the grid maximum. Nevertheless, this does not completely prevent the Trapezoidal-LS from converging at the grid boundaries in challenging cases. For instance, Figure~\ref{fig:depth_pred_vs_true} and ~\ref{fig:dur_pred_vs_true} show instances where the Trapezoidal-LS predictions saturate at the minimum and maximum grid values for depth and duration. This occurs because the transit signal is poorly defined, causing the Trapezoidal-LS to erroneously fit stellar variability instead of the true planetary event. Further extending the grid would only spread these boundary predictions into a wider range without addressing the underlying failure, producing even larger individual errors at a higher computational cost. \par

Unlike the Trapezoidal-LS, the CNN can generalize beyond these training ranges and produce estimates closer to the true value, whereas the Trapezoidal-LS is inherently bounded by its search grid. Figure~\ref{fig:dur_pred_vs_true} shows that CNN maintains consistent predictive accuracy across the entire range of transit durations, avoiding extreme failures at the boundaries. We observed a slight tendency to overestimate the duration for short-duration transits ($<$ 4 h), and even in these cases, the CNN predictions remain close to the true value. This behavior also acts as a conservative constraint that ensures the entire transit shape, particularly the ingress and egress phases, is safely contained within the predicted window rather than being clipped. The Trapezoidal-LS, in contrast, tends to underestimate the duration for short-duration transits (Figure~\ref{fig:dur_pred_vs_true}). This systematic underestimation is expected when fitting a trapezoidal profile, since the mismatch between the assumed shape and the true transit profile typically causes the fit to place the transit boundaries inside the actual ingress and egress. \par 

Table~\ref{tab:cnn_vs_ls} summarizes the performance of both methods on the 120 evaluated planets. Both methods achieve comparable median depth errors (201 ppm for the CNN and 196 ppm for the Trapezoidal-LS) but the CNN produces a substantially lower MAE (717 vs 788 ppm for depth, 1.12 vs 1.92 h for duration). This contrast between median and MAE reflects the behavior of each method: in most cases both fits recover the transit with comparable accuracy, but in a non-negligible fraction of cases the Trapezoidal-LS fails catastrophically by converging on stellar variability features or on the grid boundaries, producing errors that reach the limits of the search grid (Figure~\ref{fig:dur_pred_vs_true}). For $\Delta t$, only 2 per cent of CNN predictions have errors larger than 5 h, compared with 13 per cent for the Trapezoidal-LS baseline. At the more stringent 10 h threshold, the failure rate is 1 per cent for the CNN versus 10 per cent for the Trapezoidal-LS baseline. Because the Trapezoidal-LS fits depth, duration, and mid-transit time jointly, a failure to locate the transit also corrupts the recovered depth and duration for that same fit, even when their numerical values fall close to the truth. \par

In Figure~\ref{fig:result_confirmed_planet}, we show three representative confirmed TESS planets, comparing the CNN and Trapezoidal-LS predictions on the same LCs. As in the synthetic tests (Section~\ref{sec:baseline_comparison}), the CNN performs better than Trapezoidal-LS under high stellar variability and noise. This behavior is due to the fact that the CNN learns to distinguish these effects from the transit signal during training, whereas the Trapezoidal-LS, which does not model the out-of-transit variability, converges on stellar modulations or at the grid boundaries. Furthermore, our NN provides an uncertainty estimate for each of the three parameters together with its prediction, without additional per-target sampling or optimization. This capability is a key complementary advantage, as it explicitly flags low-confidence predictions, such as those for ambiguous shallow transits, reducing the risk of propagating unreliable parameters during data analysis. \par

\section{New Candidates sectors 27-60}
\label{sec:new_candidates}

In this section, we describe the application of our NN to new transit candidates. To achieve this, we implemented two stages: first, we utilized a DL architecture to detect new exoplanet transit candidates in sectors 27-60. Then, we applied our proposed NN to characterize these candidates, obtaining transit parameters together with their uncertainty estimates. \par

\subsection{Transit detections}
\label{sec:transit_detection}

As an initial step, we adopted the Transformer architecture proposed by \citet{salinas2025} without making any modifications to its internal structure or hyperparameter configuration. We retrained it following the same ground truth definition as \citet{salinas2025}, extended to LCs from TESS Sectors 1 to 60 to include systematic noise and instrumental trends of the more recent sectors. Positive labels correspond exclusively to confirmed and known planets from ExoFOP-TESS\footnote{\url{https://exofop.ipac.caltech.edu/tess/}}, and negative labels to EBs and non-transit signals. Unconfirmed planet candidates are excluded from the training set. Because the original architecture was optimized for FFIs at a 30-minute cadence, we binned all high-cadence LCs in our sample to match this 30-minute sampling for both the training and inference process. \par

Following the retraining of the Transformer architecture \citep{salinas2025}, we performed an inference search to identify new exoplanet candidates across Sectors 27 to 60, using the same selection criteria as \citet{salinas2025}. We excluded any candidate already listed as a TOI or a Community TESS Object of Interest (CTOI) on ExoFOP-TESS. Since every positive label in the training set corresponds to a planet already listed on ExoFOP, none of the reported candidates can belong to the positive training sample. This search yielded a total of 169 new planetary system candidates, including multi-transit, single-transit events, and multi-planet systems around the same host star. \par

\subsection{Parameter Estimation}
\label{sec:parameter_estimation}

To characterize the transit parameters of our candidates, we applied our CNN, which predicts transit depth, duration, and mid-transit time together with their associated uncertainties. For multi-transit systems, we first ran BLS to determine the orbital period, and then applied our CNN to estimate the transit parameters of each event. For single-transit candidates and for the individual transits of multi-planet systems, we applied the CNN directly to derive the transit parameters. After obtaining the transit parameters with our CNN, we convert the predicted transit depths into planetary radii as $r_p = r_\star \sqrt{d\,(c + 1)}$, where $r_\star$ is the host stellar radius, $d$ is the predicted depth, and $c$ is the background contamination of the target. Both $r_\star$ and $c$ are retrieved from the TIC via MAST using the \textsc{ASTROQUERY} package, and when the star parameters are not available, we use \textit{Gaia} mission's \citep{gaiadr3} data release 3 as a secondary source. \par

\begin{figure}
  \centering
  \includegraphics[width=0.9\linewidth]{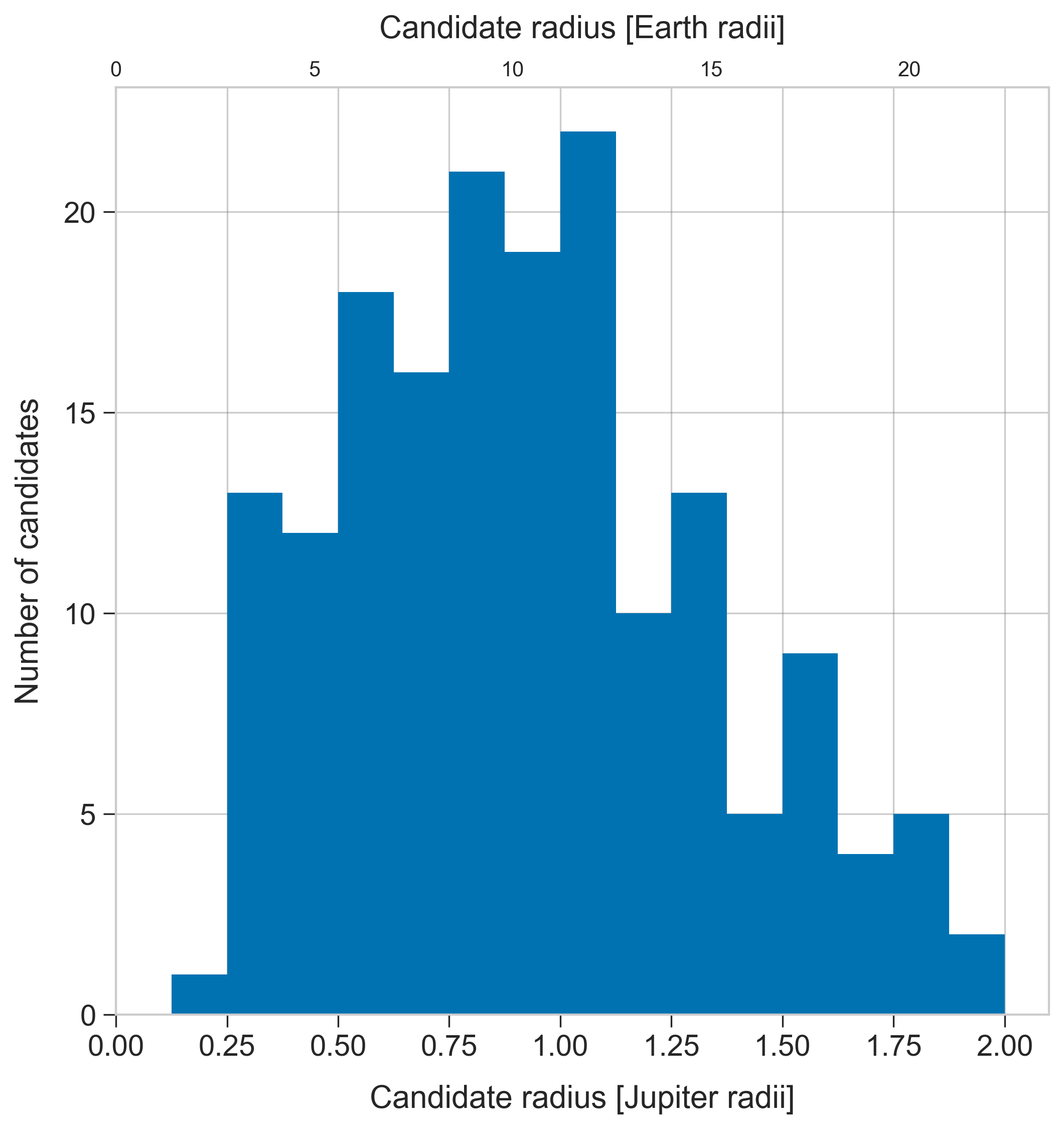}
\caption{Distribution of our candidate radii, measured in Jupiter radii.}  
    \label{fig:radius_histogram}
\end{figure}

\begin{figure}
  \centering
  \includegraphics[width=0.9\linewidth]{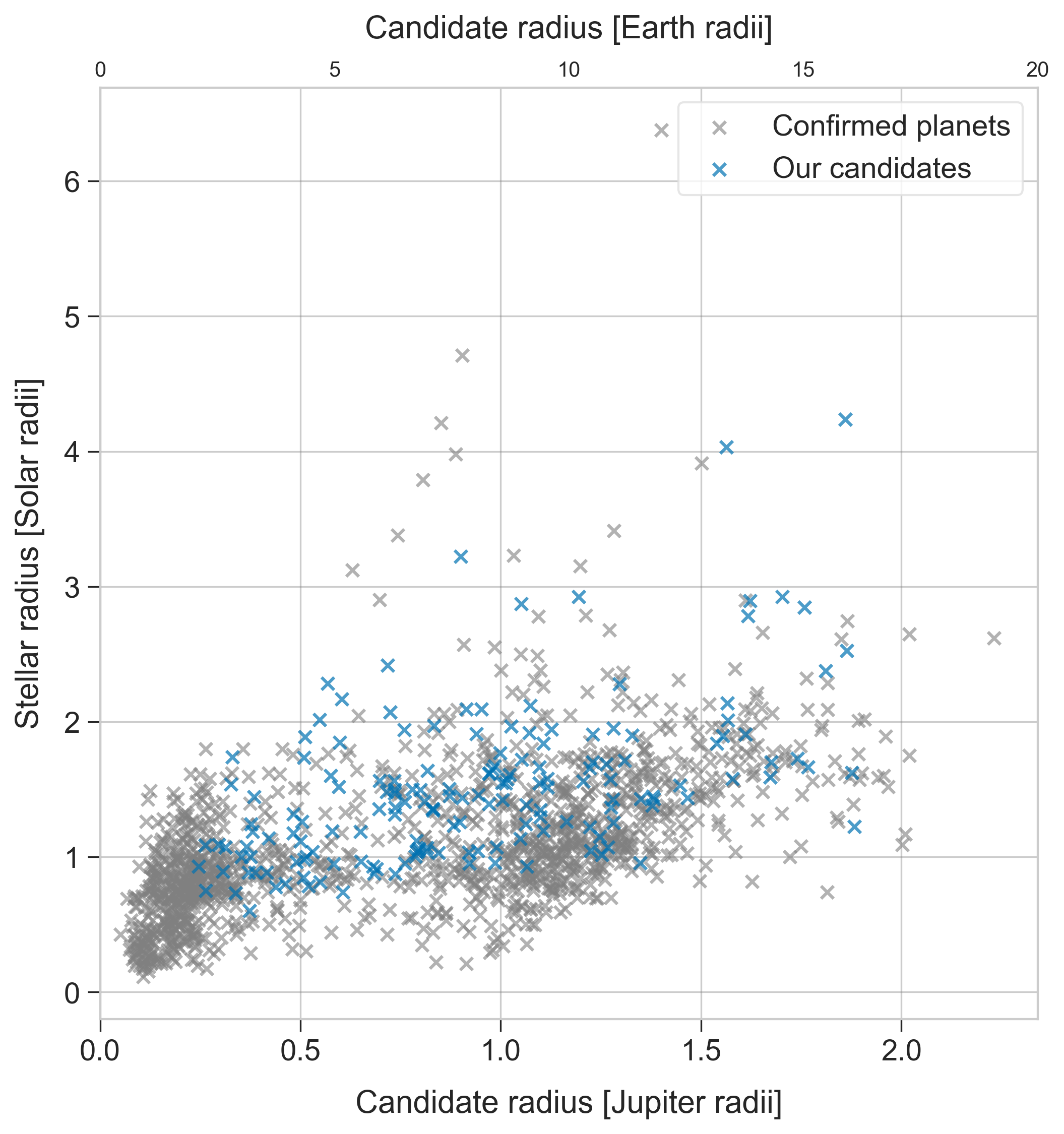}
\caption{Distribution of stellar radius and candidate radius for our candidates and for confirmed planets from ExoFOP-TESS.}  
    \label{fig:stellar_radius_vs_planet_radius}
\end{figure}

\begin{figure}
  \centering
  \includegraphics[width=0.9\linewidth]{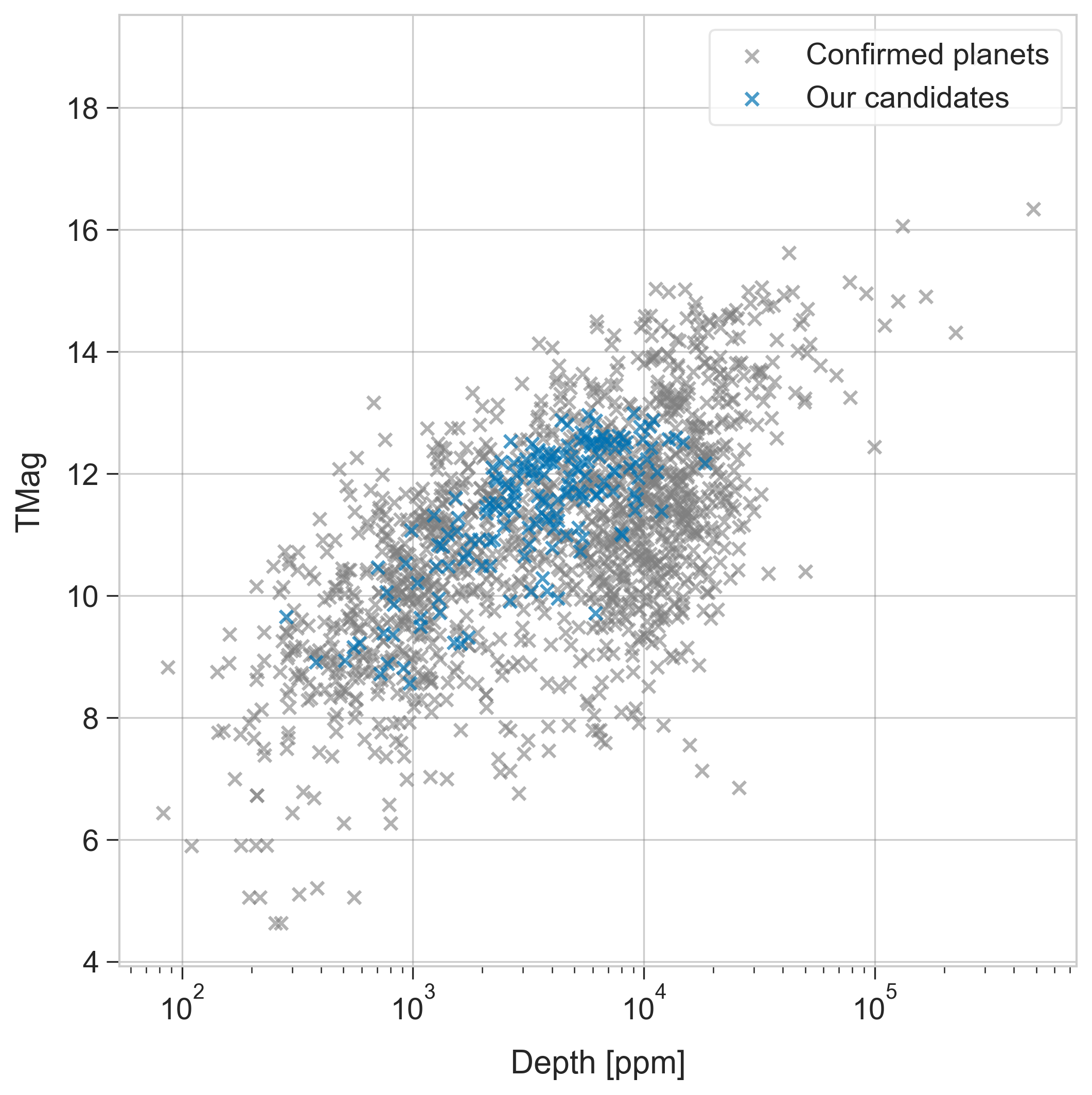}
\caption{Distribution of TESS magnitude and transit depth for our candidates and for confirmed planets from ExoFOP-TESS.}  
    \label{fig:tessmag_vs_depth}
\end{figure}

Figure~\ref{fig:radius_histogram} presents the radius distribution of our candidates, expressed in Jupiter radii $R_{\rm J}$. Most candidates are concentrated between 0.5 and 1.25 $R_{\rm J}$, with a significant peak around 1 $R_{\rm J}$. Figure~\ref{fig:stellar_radius_vs_planet_radius} shows the relationship between the candidate radius and the stellar radius of the host, and Figure~\ref{fig:tessmag_vs_depth} shows the relationship between TESS magnitude and transit depth, in both cases together compared to confirmed planets. Our candidates are concentrated around stars between 0.8 and 2 solar radii and TESS magnitudes between 10 and 13, with transit depths of $10^3$ to $10^4$ ppm. Only a small fraction of our candidates come from small, faint stars, which is precisely the region where confirmed sub-Neptunes are concentrated. \par

\begin{figure}
  \centering
  \includegraphics[width=0.9\linewidth]{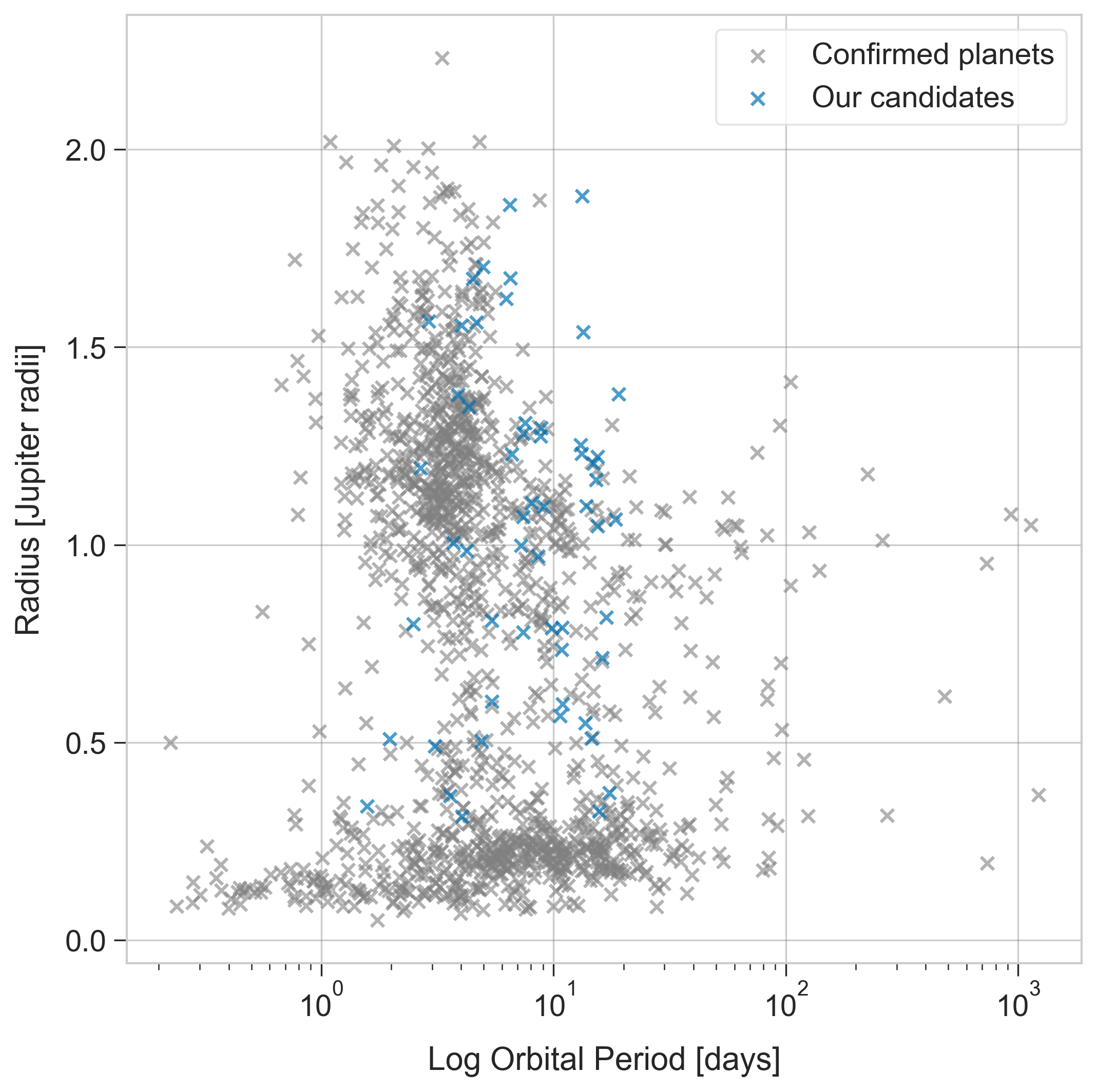}
\caption{Distribution of planetary radius and log orbital period for our candidates and for confirmed planets from ExoFOP-TESS}  
    \label{fig:radius_vs_period}
\end{figure}

For multi-transit candidates, we compare our candidates with the confirmed planet population in terms of radius and orbital period. Figure~\ref{fig:radius_vs_period} shows that most of our candidates have periods below 20 days and radii between 0.5 and 1.5 $R_{\rm J}$, overlapping the region already populated by confirmed giant planets at short periods. \par

The complete list of candidates is reported in Table~\ref{tab:single_candidates}, Table~\ref{tab:multi_transit_candidates}, and Table~\ref{tab:multi_planet_candidates}, which list the single transit, multi-transit, and multi-planet detections together with the CNN estimates of depth, duration, and $t_0$ and their predicted uncertainties. Overall, our CNN provides transit parameters and uncertainties for all 169 candidates. Most importantly, it characterizes the single transit and the individual transits of multi-planet systems. These events cannot be handled by phase-folding techniques.

\section{Discussion}
\label{sec:discussion}

Our network learns to predict transit parameters directly from the observed flux, handling stellar variability without requiring prior detrending or per-target noise modeling. The main advantage of our CNN over the Trapezoidal-LS baseline is its robustness rather than in its typical accuracy. On confirmed TESS planets, both methods achieve comparable median errors, but the Trapezoidal-LS baseline fails to locate the transit in 10 per cent of cases, compared with only 1 per cent for our CNN. This difference arises because the Trapezoidal-LS fits only the local flux without modeling the out-of-transit variability, which causes the fit to converge onto stellar variability features or saturate at search grid boundaries when photometric noise is high. In contrast, the learned convolutional filters of our CNN use the out-of-transit flux as context to distinguish local transit shape from low-frequency stellar modulations. These failures would propagate incorrect parameters into follow-up prioritization. In terms of computational cost, our CNN processes 20,000 LCs in 0.374 s (0.019 ms per LC), roughly 8 times faster than the Trapezoidal-LS baseline, which takes 3.0 s (0.15 ms per LC). In addition, the computational cost of the BLS-based fitting methods depends on the parameter grid searched, whereas the computational cost of our CNN is constant per candidate. \par

Despite our CNN's accuracy in characterizing transit parameters, our network still requires an approximate transit epoch $\hat{t_0}$ as input. Although the network does not require this epoch to be precisely centered within the window, since the soft-argmax head corrects for offsets of up to several hours, it cannot locate a transit within an arbitrarily long LC on its own. If the network were instead able to recover the mid-transit time directly from the full LC, it would function not only as a characterization tool but also as a detection method. However, this possibility goes beyond the scope of this work, which focuses specifically on the characterization of transit candidates already identified. In practice, this detection step can be performed by networks that operate on the full LC such as the Transformer of \citet{salinas2025}, which we combine with our CNN in Section~\ref{sec:new_candidates}, or the pipeline of \citet{olmschenk2021identifying}. \par

Although our synthetic data incorporate multiple realistic sources of noise and stellar variability, they cannot reproduce the full distribution of TESS systematics, and therefore performance on real data remains the most relevant test of generalization. While our data simulations include slow photometric drifts, correlated noise, stellar variability, and isolated outliers (Section~\ref{sec:data_simulation}), they do not reproduce some TESS-specific systematics. For example, scattered light from the Earth and Moon enters the cameras near the perigee of the TESS orbit. Other examples are flux discontinuities and short-term flux ramps associated with momentum dumps \citep{wong2020systematic}, and the data gaps between the two orbits of each sector. Despite these differences, our CNN fails to locate the transit in only 2 of the 120 confirmed planets, whose LCs contain these systematics. In this context, the results on 120 confirmed planets (Section~\ref{sec:validation_know_planets}) show that the representation learned from synthetic data transfers to real observations, and that the robustness of the CNN to stellar variability is preserved. Since the uncertainty calibration is also learned on synthetic data, the reported coverage applies to the synthetic distribution, and its application to real LCs would benefit from further validation. In addition, the predicted uncertainties are learned from the conditions represented in our simulations, and may therefore be less representative for inputs that differ substantially from them, such as systematics not included in our training set. \par

\begin{figure*}
  \centering
  \includegraphics[width=0.9\linewidth]{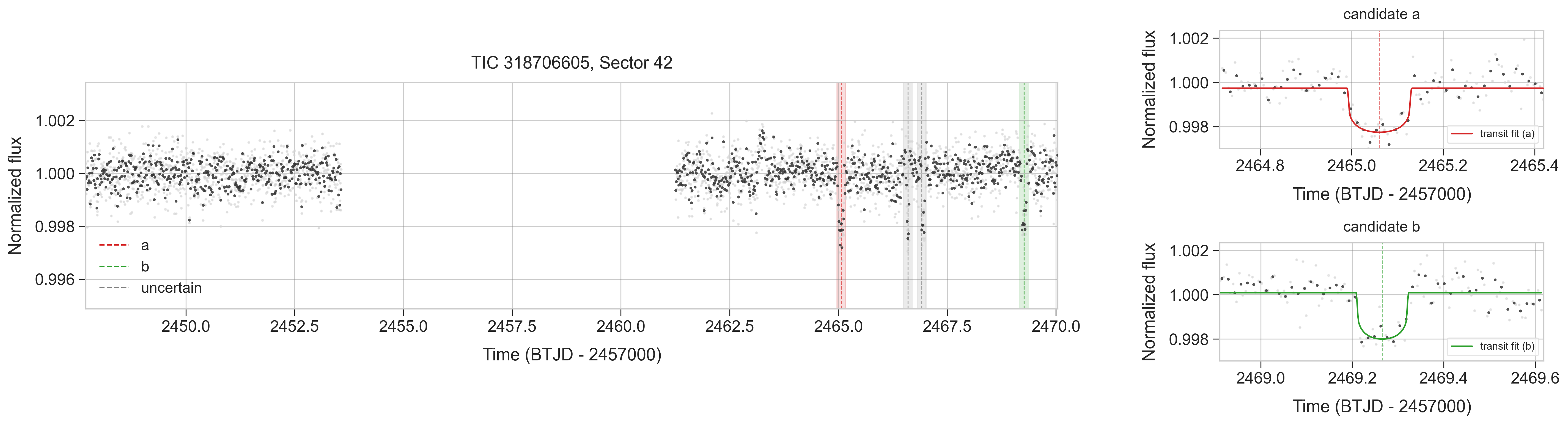}
\caption{LC of TIC 318706605 in Sector 42, showing two transit events from different planet candidates in the same system, and two possible transits (uncertain) of lower confidence. The right panels show a zoom of each candidate transit with the batman transit model computed from the CNN predictions.}  
    \label{fig:TIC318706605_lightcurves}
\end{figure*}

Beyond the validation on confirmed planets, we applied the complete pipeline to TESS Sectors 27 to 60 using the detection model of \citet{salinas2025} to identify new candidates, and then our CNN to estimate their physical parameters. Among the 169 new candidates characterized, we highlight three systems that show the ability of our CNN to characterize individual transit events that cannot be recovered through phase-folding. TIC 318706605, observed in Sector 42 (Figure~\ref{fig:TIC318706605_lightcurves}), shows two distinct single transit events from different planet candidates in the same system with similar depths ($\sim$2160 and $\sim$1990 ppm). Despite the small difference in depth, we treat them as separate planets rather than a multi-transit candidate, since a common period matching their separation would produce additional transits within the observed sector, which are absent. Figure~\ref{fig:TIC318706605_lightcurves} also shows two additional dips near BTJD 2466.6 and 2466.9, but they cannot be transits of candidates a or b, since the corresponding periods would also produce additional transits within the observed sector. The centroid and background time series show no variations during these dips, which disfavors a contaminating source, and they may correspond to additional exoplanet candidates. Confirming them and constraining their periods requires the analysis of additional sectors. The nearest neighbor, TIC 318706606 ($4.85''$, $\sim$0.23 px, $\Delta T \approx 5.05$ mag), where $T$ is the magnitude of the TESS-band, is fully blended and cannot be excluded as the transit source based solely on its blended flux \citep{seager2003unique}. However, this scenario is disfavored by the absence of a secondary eclipse and by the two independent signals in the same aperture, which reduce the false positive probability \citep{lissauer2012almost}. The target also has a low \textit{Gaia} radial velocity (RV) error of $0.382$ km s$^{-1}$. A short-period stellar companion would inflate this value by varying the star's velocity between \textit{Gaia} observations, but its low level is consistent with the absence of such a companion \citep{katz2023gaia}. More studies and analysis are required, including follow-up observations, to exclude a false positive origin. \par

\begin{figure*}
  \centering
  \includegraphics[width=0.9\linewidth]{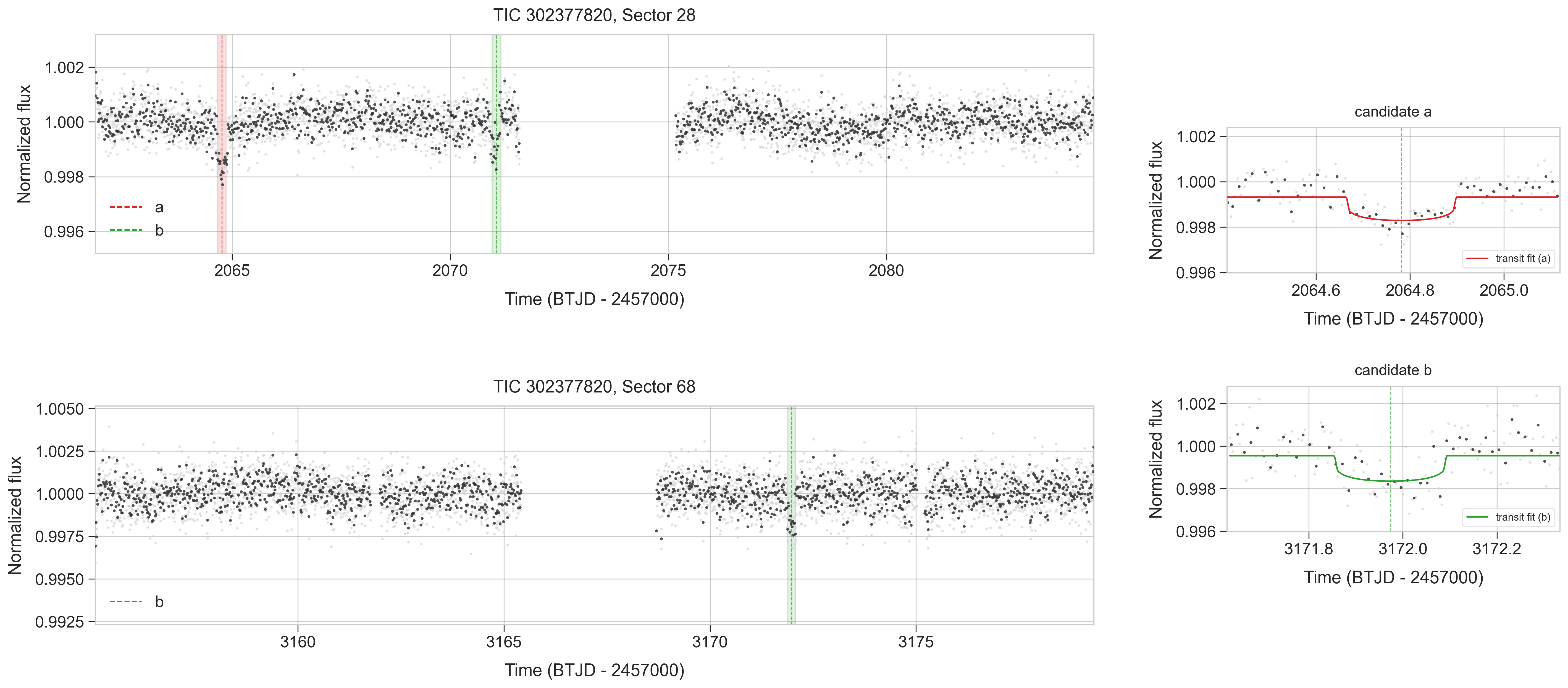}
\caption{LC of TIC 302377820 in Sector 28 (top) and Sector 68 (bottom), showing three transit events from two different planet candidates in the same system. The right panels show a zoom of the transit of candidate a in Sector 28 and of candidate b in Sector 68 with a transit model based using the CNN predictions.}  
    \label{fig:TIC302377820_lightcurves}
\end{figure*}

TIC 302377820 shows two candidates detected in Sector 28 (Figure~\ref{fig:TIC302377820_lightcurves}), with depths of $\sim$1420 and $\sim$1255~ppm and radii of $R_{\rm p} \approx$ 0.38 and 0.35 $R_{\rm J}$. We additionally inspected Sector 68, where a further transit appears with a depth similar to that of candidate ``b''. This, together with the presence of two independent signals in the same aperture, reduces the probability of a false positive origin \citep{lissauer2012almost}. As with the previously described candidate, the target also has a low RV error of $0.256$ km s$^{-1} $ consistent with the absence of a short-period stellar companion, though not conclusive \citep{katz2023gaia}. The only relevant contaminant, TIC 302377823 ($23.1''$, $\sim$1.10 px, $\Delta T \approx 5.52$ mag), is an M dwarf ($T_{\rm eff} \approx 3530$ K) which cannot be excluded as the transit source from its blended flux alone \citep{seager2003unique}. \par

\begin{figure*}
  \centering
  \includegraphics[width=0.9\linewidth]{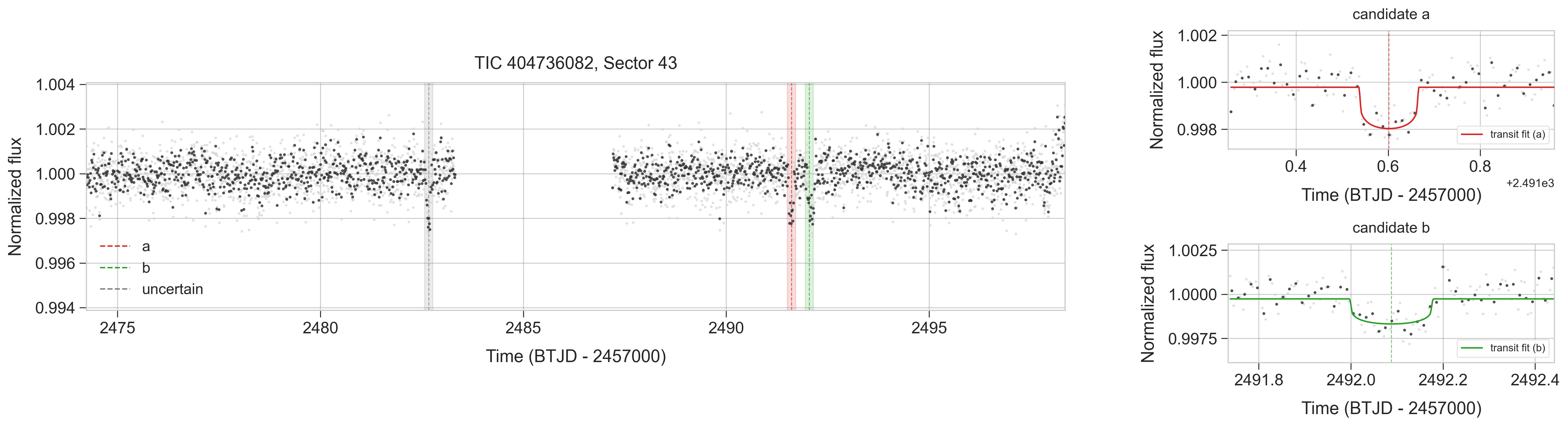}
\caption{LC of TIC 404736082 in Sector 43, showing two transit events from two different planet candidates, and a possible third transit (uncertain) of lower confidence. The right panels show a zoom of the transits of candidates a and b, with a batman model using the CNN predictions.}  
    \label{fig:TIC404736082_lightcurves}
\end{figure*}

TIC 404736082 shows two transit candidates in Sector 43 (Figure~\ref{fig:TIC404736082_lightcurves}), separated by $\sim$10.5 h, with depths of $\sim$2230 and $\sim$1780 ppm and radii of $R_{\rm p} \approx$ 0.42 and 0.37 $R_{\rm J}$, and with an RV error of $0.550$ km~s$^{-1}$. The absence of periodic transits at this spacing excludes a single short-period object. We therefore treat them as separate planets. An additional shallower event near 2482.67 (BTJD - 2457000) shows a more V-shaped profile, and we consider it of lower confidence. Other transits are present in Sector 70, although their association with the individual candidates remains unclear and requires further analysis. The nearest neighbor, TIC 404736084 ($5.07''$, $\sim$0.24 px, $\Delta T \approx 3.50$ mag), is blended and cannot be excluded as the transit source \citep{seager2003unique}. However, the absence of a secondary eclipse disfavors a blended EB in its position. As with the previous systems, these scenarios can only be excluded with further analysis and follow-up. \par

\section{Conclusions}
\label{sec:conclusions}

This work presented a CNN that predicts transit depth, duration, and mid-transit time, together with calibrated uncertainties, directly from single-transit LCs. Our network is trained on synthetic LCs that include stellar variability and noise, and it requires neither prior detrending nor per-target fitting. This addresses a fundamental limitation of single transit characterization: methods robust to stellar variability typically require independent fitting for each candidate, while faster BLS approaches are prone to absorbing stellar variability into the fitted parameters. \par

Our proposed model recovers transit depth with a MAE of 170 ppm, duration with a MAE of 0.88 h, and mid-transit time with a MAE of 1.53 h on our synthetic test set, with coverage close to 68 and 95 per cent levels, indicating well-calibrated uncertainties. In our sensitivity tests, performance declines gradually with increasing noise and stellar variability, most notably in the mid-transit time, while depth and duration remain comparatively stable. On synthetic data with noise and variability varying simultaneously, the CNN achieves a lower failure rate than the Trapezoidal-LS baseline across the SNR range, and a failure rate two to three times lower across the full range of stellar variability amplitudes, with the same parameter search ranges for both methods. \par

The main advantage of our CNN is its robustness in the most difficult cases. On 120 confirmed TESS planets, the CNN and the Trapezoidal-LS baseline achieve comparable median depth errors, but the CNN produces a substantially lower MAE for depth, duration, and mid-transit time. At the 5 h threshold, only 2 per cent of CNN predictions fail to recover the transit epoch, compared with 13 per cent for the Trapezoidal-LS baseline. This gap arises because the Trapezoidal-LS jointly fits depth, duration, and mid-transit time from the same profile, and a failure to locate the transit also corrupts the recovered depth and duration for that same fit. In such cases, the resulting depth or duration estimate may still appear numerically close to the true value. However, this reflects convergence on a stellar variability feature that happens to mimic the transit signal under high noise or variability, rather than a correct identification of the transit itself. \par

Applied to 169 new candidates identified across TESS Sectors 27 to 60, our CNN provided transit parameters and uncertainty estimates for single transit events, multi-transit systems, and individual transits within multi-planet systems. Single transits and the individual transits of multi-planet systems cannot be phase-folded, and characterizing them with robust per-target fitting does not scale to large candidate samples. In addition, the computational cost of our CNN is constant per candidate, whereas that of BLS-based fitting depends on the parameter grid searched. This constant cost is an advantage for the growing volume of single transit detections expected from current and future surveys. \par

Our future work aims to build a more complete pipeline for transit analysis, integrating detection and characterization. Since this would require substantially larger training sets, a promising strategy is transfer learning, where a network pre-trained on large volumes of unlabeled LCs first learns the stellar variability and systematics directly from observational data and is then fine-tuned for transit detection and characterization. Beyond increasing the amount of training data, this approach would reduce the dependence on the assumptions of our synthetic simulations, which cannot reproduce the full distribution of TESS systematics. As shown in previous work \citep{olmschenk2021identifying, salinas2025}, rather than applying detrending, NNs can learn a representation of the LC that helps to distinguish transits from background modulations. This representation could then be adapted with limited additional training to other tasks, such as false positive vetting or the characterization of stellar variability. It could also be applied to TESS observations at other cadences and to upcoming missions such as the Nancy Grace Roman Space Telescope \citep{spergel2015wide} and the PLAnetary Transits and Oscillations of stars telescope \citep[PLATO;][]{rauer2016plato}, which will present new challenges for both the detection and characterization of exoplanets. \par

\section*{Acknowledgements}

HS and RB acknowledge the support of the FONDECYT project 1241963. RB also acknowledges support from the ANID--Millennium Science Initiative--ICN12\_009. M.S.-F. acknowledges support from the Heising-Simons Foundation through grant 2026-6822 (Unsolved Problems in Astrophysics Mentorship Network). The material is based upon work supported by NASA under award number 80GSFC24M0006.

We acknowledge the use of public information of the TESS mission. Funding for the TESS mission is provided by the NASA Explorer Program. Thanks to the availability of data from TESS, we were able to examine and display our analysis of transit signals.

This research has made use of the Exoplanet Follow-up Observation Program (ExoFOP; doi: 10.26134/ExoFOP5) website, which is operated by the California Institute of Technology, under contract with the National Aeronautics and Space Administration under the Exoplanet Exploration Program.

\section*{Data Availability}

All the data referenced here and that were used is public information about the TESS mission from MAST, as well as data from the European Space Agency (ESA) mission \textit{Gaia}.



\bibliographystyle{mnras}
\bibliography{bibl} 

\begin{thebibliography}{}
\makeatletter
\relax
\def\mn@urlcharsother{\let\do\@makeother \do\$\do\&\do\#\do\^\do\_\do\%\do\~}
\def\mn@doi{\begingroup\mn@urlcharsother \@ifnextchar [ {\mn@doi@} {\mn@doi@[]}}
\def\mn@doi@[#1]#2{\def\@tempa{#1}\ifx\@tempa\@empty \href {http://dx.doi.org/#2} {doi:#2}\else \href {http://dx.doi.org/#2} {#1}\fi \endgroup}
\def\mn@eprint#1#2{\mn@eprint@#1:#2::\@nil}
\def\mn@eprint@arXiv#1{\href {http://arxiv.org/abs/#1} {{\tt arXiv:#1}}}
\def\mn@eprint@dblp#1{\href {http://dblp.uni-trier.de/rec/bibtex/#1.xml} {dblp:#1}}
\def\mn@eprint@#1:#2:#3:#4\@nil{\def\@tempa {#1}\def\@tempb {#2}\def\@tempc {#3}\ifx \@tempc \@empty \let \@tempc \@tempb \let \@tempb \@tempa \fi \ifx \@tempb \@empty \def\@tempb {arXiv}\fi \@ifundefined {mn@eprint@\@tempb}{\@tempb:\@tempc}{\expandafter \expandafter \csname mn@eprint@\@tempb\endcsname \expandafter{\@tempc}}}

\bibitem[\protect\citeauthoryear{Aigrain, Hodgkin, Irwin, Lewis  \& Roberts}{Aigrain et~al.}{2015}]{aigrain2015precise}
Aigrain S.,  Hodgkin S.~T.,  Irwin M.~J.,  Lewis J.~R.,   Roberts S.~J.,  2015, \mn@doi [Monthly Notices of the Royal Astronomical Society] {10.1093/mnras/stu2638}, 447, 2880

\bibitem[\protect\citeauthoryear{Angus, Morton, Aigrain, Foreman-Mackey  \& Rajpaul}{Angus et~al.}{2018}]{angus2018inferring}
Angus R.,  Morton T.,  Aigrain S.,  Foreman-Mackey D.,   Rajpaul V.,  2018, \mn@doi [Monthly Notices of the Royal Astronomical Society] {10.1093/mnras/stx2109}, 474, 2094

\bibitem[\protect\citeauthoryear{Bengio}{Bengio}{2012}]{bengio2012practical}
Bengio Y.,  2012, in Montavon G.,  Orr G.~B.,   M{\"u}ller K.-R.,  eds, Lecture Notes in Computer Science, Vol.~7700, Neural Networks: Tricks of the Trade, Second Edition.
Springer, Berlin, Heidelberg, pp 437--478, \mn@doi{10.1007/978-3-642-35289-8_26}

\bibitem[\protect\citeauthoryear{Bieryla et~al.,}{Bieryla et~al.}{2026}]{bieryla2026toi}
Bieryla A.,  et~al., 2026, \mn@doi [The Astronomical Journal] {10.3847/1538-3881/ae3d9f}, 171, 203

\bibitem[\protect\citeauthoryear{Borucki et~al.,}{Borucki et~al.}{2010}]{borucki2010kepler}
Borucki W.~J.,  et~al., 2010, Science, 327, 977

\bibitem[\protect\citeauthoryear{Brahm et~al.,}{Brahm et~al.}{2023}]{brahm2023three}
Brahm R.,  et~al., 2023, \mn@doi [The Astronomical Journal] {10.3847/1538-3881/accadd}, 165, 227

\bibitem[\protect\citeauthoryear{Caceres, Feigelson, Babu, Bahamonde, Christen, Bertin, Meza  \& Cur{\'e}}{Caceres et~al.}{2019}]{caceres2019autoregressive}
Caceres G.~A.,  Feigelson E.~D.,  Babu G.~J.,  Bahamonde N.,  Christen A.,  Bertin K.,  Meza C.,   Cur{\'e} M.,  2019, \mn@doi [The Astronomical Journal] {10.3847/1538-3881/ab26b8}, 158, 57

\bibitem[\protect\citeauthoryear{Dalba et~al.,}{Dalba et~al.}{2022}]{dalba2022tess}
Dalba P.~A.,  et~al., 2022, \mn@doi [The Astronomical Journal] {10.3847/1538-3881/ac415b}, 163, 61

\bibitem[\protect\citeauthoryear{Dawson \& Johnson}{Dawson \& Johnson}{2018}]{dawson2018origins}
Dawson R.~I.,  Johnson J.~A.,  2018, \mn@doi [Annual Review of Astronomy and Astrophysics] {10.1146/annurev-astro-081817-051853}, 56, 175

\bibitem[\protect\citeauthoryear{Eschen \& Kunimoto}{Eschen \& Kunimoto}{2024}]{eschen2024nine}
Eschen Y. N.~E.,  Kunimoto M.,  2024, \mn@doi [Monthly Notices of the Royal Astronomical Society] {10.1093/mnras/stae1496}, 531, 5053

\bibitem[\protect\citeauthoryear{Essack et~al.,}{Essack et~al.}{2025}]{essack2025giant}
Essack Z.,  et~al., 2025, \mn@doi [The Astronomical Journal] {10.3847/1538-3881/add88b}, 170, 41

\bibitem[\protect\citeauthoryear{Foreman-Mackey, Morton, Hogg, Agol  \& Sch{\"o}lkopf}{Foreman-Mackey et~al.}{2016}]{foreman2016population}
Foreman-Mackey D.,  Morton T.~D.,  Hogg D.~W.,  Agol E.,   Sch{\"o}lkopf B.,  2016, \mn@doi [The Astronomical Journal] {10.3847/0004-6256/152/6/206}, 152, 206

\bibitem[\protect\citeauthoryear{Foreman-Mackey, Agol, Ambikasaran  \& Angus}{Foreman-Mackey et~al.}{2017}]{foreman2017fast}
Foreman-Mackey D.,  Agol E.,  Ambikasaran S.,   Angus R.,  2017, \mn@doi [The Astronomical Journal] {10.3847/1538-3881/aa9332}, 154, 220

\bibitem[\protect\citeauthoryear{Fressin et~al.,}{Fressin et~al.}{2013}]{fressin2013false}
Fressin F.,  et~al., 2013, \mn@doi [The Astrophysical Journal] {10.1088/0004-637X/766/2/81}, 766, 81

\bibitem[\protect\citeauthoryear{{Gaia Collaboration}, Vallenari, Brown  \& et al.}{{Gaia Collaboration} et~al.}{2023}]{gaiadr3}
{Gaia Collaboration} Vallenari A.,  Brown A. G.~A.,   et al. 2023, \mn@doi [Astronomy \& Astrophysics] {10.1051/0004-6361/202243940}, 674, A1

\bibitem[\protect\citeauthoryear{Gill et~al.,}{Gill et~al.}{2020}]{gill2020ngts}
Gill S.,  et~al., 2020, \mn@doi [Monthly Notices of the Royal Astronomical Society] {10.1093/mnras/stz3212}, 491, 1548

\bibitem[\protect\citeauthoryear{G{\"u}nther \& Daylan}{G{\"u}nther \& Daylan}{2021}]{gunther2021allesfitter}
G{\"u}nther M.~N.,  Daylan T.,  2021, The Astrophysical Journal Supplement Series, 254, 13

\bibitem[\protect\citeauthoryear{Guo, Pleiss, Sun  \& Weinberger}{Guo et~al.}{2017}]{guo2017calibration}
Guo C.,  Pleiss G.,  Sun Y.,   Weinberger K.~Q.,  2017, in Proceedings of the 34th International Conference on Machine Learning. pp 1321--1330 (\mn@eprint {arXiv} {1706.04599})

\bibitem[\protect\citeauthoryear{Hartman et~al.,}{Hartman et~al.}{2019}]{hartman2019hats}
Hartman J.~D.,  et~al., 2019, \mn@doi [The Astronomical Journal] {10.3847/1538-3881/aaf8b6}, 157, 55

\bibitem[\protect\citeauthoryear{Hawthorn et~al.,}{Hawthorn et~al.}{2024}]{hawthorn2024tess}
Hawthorn F.,  et~al., 2024, \mn@doi [Monthly Notices of the Royal Astronomical Society] {10.1093/mnras/stad3783}, 528, 1841

\bibitem[\protect\citeauthoryear{Haywood et~al.,}{Haywood et~al.}{2014}]{haywood2014planets}
Haywood R.~D.,  et~al., 2014, \mn@doi [Monthly Notices of the Royal Astronomical Society] {10.1093/mnras/stu1320}, 443, 2517

\bibitem[\protect\citeauthoryear{Hendrycks \& Gimpel}{Hendrycks \& Gimpel}{2016}]{hendrycks2016gaussian}
Hendrycks D.,  Gimpel K.,  2016, \mn@doi [arXiv e-prints] {10.48550/arXiv.1606.08415}, p. arXiv:1606.08415

\bibitem[\protect\citeauthoryear{Hippke \& Heller}{Hippke \& Heller}{2019}]{hippke2019optimized}
Hippke M.,  Heller R.,  2019, \mn@doi [Astronomy \& Astrophysics] {10.1051/0004-6361/201834672}, 623, A39

\bibitem[\protect\citeauthoryear{Incha et~al.,}{Incha et~al.}{2023}]{incha2023kepler}
Incha E.,  et~al., 2023, \mn@doi [Monthly Notices of the Royal Astronomical Society] {10.1093/mnras/stad1049}, 523, 474

\bibitem[\protect\citeauthoryear{Ioffe \& Szegedy}{Ioffe \& Szegedy}{2015}]{ioffe2015batch}
Ioffe S.,  Szegedy C.,  2015, in Proceedings of the 32nd International Conference on Machine Learning. pp 448--456 (\mn@eprint {arXiv} {1502.03167})

\bibitem[\protect\citeauthoryear{Jenkins et~al.,}{Jenkins et~al.}{2010}]{jenkins2010overview}
Jenkins J.~M.,  et~al., 2010, \mn@doi [The Astrophysical Journal Letters] {10.1088/2041-8205/713/2/L87}, 713, L87

\bibitem[\protect\citeauthoryear{Katz et~al.,}{Katz et~al.}{2023}]{katz2023gaia}
Katz D.,  et~al., 2023, \mn@doi [Astronomy \& Astrophysics] {10.1051/0004-6361/202244220}, 674, A5

\bibitem[\protect\citeauthoryear{Kendall \& Gal}{Kendall \& Gal}{2017}]{kendall2017uncertainties}
Kendall A.,  Gal Y.,  2017, in Advances in Neural Information Processing Systems 30 (NIPS 2017). pp 5574--5584, \mn@doi{10.48550/arXiv.1703.04977}

\bibitem[\protect\citeauthoryear{Kendall et~al.,}{Kendall et~al.}{2026}]{kendall202643}
Kendall A.,  et~al., 2026, \mn@doi [Monthly Notices of the Royal Astronomical Society] {10.48550/arXiv.2512.07716}, 547, staf2189

\bibitem[\protect\citeauthoryear{Kingma \& Ba}{Kingma \& Ba}{2015}]{kingma2014adam}
Kingma D.~P.,  Ba J.,  2015, in International Conference on Learning Representations (ICLR). , \mn@doi{10.48550/arXiv.1412.6980}

\bibitem[\protect\citeauthoryear{Kipping}{Kipping}{2008}]{kipping2008transiting}
Kipping D.~M.,  2008, \mn@doi [Monthly Notices of the Royal Astronomical Society] {10.1111/j.1365-2966.2008.13658.x}, 389, 1383

\bibitem[\protect\citeauthoryear{Kipping}{Kipping}{2013}]{kipping2013efficient}
Kipping D.~M.,  2013, Monthly Notices of the Royal Astronomical Society, 435, 2152

\bibitem[\protect\citeauthoryear{Kipping}{Kipping}{2023}]{kipping2023snr}
Kipping D.,  2023, \mn@doi [Monthly Notices of the Royal Astronomical Society] {10.48550/arXiv.2305.06790}, 523, 1182

\bibitem[\protect\citeauthoryear{Kov{\'a}cs, Zucker  \& Mazeh}{Kov{\'a}cs et~al.}{2002}]{kovacs2002box}
Kov{\'a}cs G.,  Zucker S.,   Mazeh T.,  2002, \mn@doi [Astronomy \& Astrophysics] {10.1051/0004-6361:20020802}, 391, 369

\bibitem[\protect\citeauthoryear{Kreidberg}{Kreidberg}{2015}]{kreidberg2015batman}
Kreidberg L.,  2015, \mn@doi [Publications of the Astronomical Society of the Pacific] {10.1086/683602}, 127, 1161

\bibitem[\protect\citeauthoryear{Krizhevsky, Sutskever  \& Hinton}{Krizhevsky et~al.}{2012}]{krizhevsky2012imagenet}
Krizhevsky A.,  Sutskever I.,   Hinton G.~E.,  2012, in Advances in Neural Information Processing Systems 25 (NIPS 2012). pp 1097--1105

\bibitem[\protect\citeauthoryear{Kuleshov, Fenner  \& Ermon}{Kuleshov et~al.}{2018}]{kuleshov2018accurate}
Kuleshov V.,  Fenner N.,   Ermon S.,  2018, in Proceedings of the 35th International Conference on Machine Learning. pp 2796--2804, \mn@doi{10.48550/arXiv.1807.00263}

\bibitem[\protect\citeauthoryear{Kunimoto et~al.,}{Kunimoto et~al.}{2022}]{kunimoto2022tess}
Kunimoto M.,  et~al., 2022, \mn@doi [The Astrophysical Journal Supplement Series] {10.3847/1538-4365/ac5688}, 259, 33

\bibitem[\protect\citeauthoryear{Lakshminarayanan, Pritzel  \& Blundell}{Lakshminarayanan et~al.}{2017}]{lakshminarayanan2017simple}
Lakshminarayanan B.,  Pritzel A.,   Blundell C.,  2017, in Advances in Neural Information Processing Systems 30 (NIPS 2017). pp 6402--6413 (\mn@eprint {arXiv} {1612.01474})

\bibitem[\protect\citeauthoryear{LeCun, Bengio  \& Hinton}{LeCun et~al.}{2015}]{lecun2015deep}
LeCun Y.,  Bengio Y.,   Hinton G.,  2015, \mn@doi [Nature] {10.1038/nature14539}, 521, 436

\bibitem[\protect\citeauthoryear{Li, Tenenbaum, Twicken, Burke, Jenkins, Quintana, Rowe  \& Seader}{Li et~al.}{2019}]{li2019kepler}
Li J.,  Tenenbaum P.,  Twicken J.~D.,  Burke C.~J.,  Jenkins J.~M.,  Quintana E.~V.,  Rowe J.~F.,   Seader S.~E.,  2019, Publications of the Astronomical Society of the Pacific, 131, 024506

\bibitem[\protect\citeauthoryear{Lissauer et~al.,}{Lissauer et~al.}{2012}]{lissauer2012almost}
Lissauer J.~J.,  et~al., 2012, \mn@doi [The Astrophysical Journal] {10.1088/0004-637X/750/2/112}, 750, 112

\bibitem[\protect\citeauthoryear{Loshchilov \& Hutter}{Loshchilov \& Hutter}{2019}]{loshchilov2017decoupled}
Loshchilov I.,  Hutter F.,  2019, in International Conference on Learning Representations (ICLR). , \mn@doi{10.48550/arXiv.1711.05101}

\bibitem[\protect\citeauthoryear{Magliano et~al.,}{Magliano et~al.}{2024}]{magliano2024plato}
Magliano C.,  et~al., 2024, \mn@doi [Monthly Notices of the Royal Astronomical Society] {10.1093/mnras/stae210}, 528, 2851

\bibitem[\protect\citeauthoryear{Mandel \& Agol}{Mandel \& Agol}{2002}]{mandel2002analytic}
Mandel K.,  Agol E.,  2002, \mn@doi [The Astrophysical Journal Letters] {10.1086/345520}, 580, L171

\bibitem[\protect\citeauthoryear{Mann et~al.,}{Mann et~al.}{2023}]{mann2023giant}
Mann C.~R.,  et~al., 2023, \mn@doi [The Astronomical Journal] {10.3847/1538-3881/ad00bc}, 166, 239

\bibitem[\protect\citeauthoryear{McQuillan, Mazeh  \& Aigrain}{McQuillan et~al.}{2014}]{mcquillan2014rotation}
McQuillan A.,  Mazeh T.,   Aigrain S.,  2014, \mn@doi [The Astrophysical Journal Supplement Series] {10.1088/0067-0049/211/2/24}, 211, 24

\bibitem[\protect\citeauthoryear{Mullally et~al.,}{Mullally et~al.}{2015}]{mullally2015planetary}
Mullally F.,  et~al., 2015, \mn@doi [The Astrophysical Journal Supplement Series] {10.1088/0067-0049/217/2/31}, 217, 31

\bibitem[\protect\citeauthoryear{Nelder \& Mead}{Nelder \& Mead}{1965}]{nelder1965simplex}
Nelder J.~A.,  Mead R.,  1965, \mn@doi [The Computer Journal] {10.1093/comjnl/7.4.308}, 7, 308

\bibitem[\protect\citeauthoryear{Nix \& Weigend}{Nix \& Weigend}{1994}]{nix1994estimating}
Nix D.~A.,  Weigend A.~S.,  1994, in Proceedings of 1994 IEEE International Conference on Neural Networks (ICNN'94). pp 55--60, \mn@doi{10.1109/ICNN.1994.374138}

\bibitem[\protect\citeauthoryear{Olmschenk et~al.,}{Olmschenk et~al.}{2021}]{olmschenk2021identifying}
Olmschenk G.,  et~al., 2021, The Astronomical Journal, 161, 273

\bibitem[\protect\citeauthoryear{Osborn}{Osborn}{2022}]{osborn2022monotools}
Osborn H.~P.,  2022, Astrophysics Source Code Library, record ascl:2204.020

\bibitem[\protect\citeauthoryear{Osborn et~al.,}{Osborn et~al.}{2016}]{osborn2016single}
Osborn H.~P.,  et~al., 2016, \mn@doi [Monthly Notices of the Royal Astronomical Society] {10.1093/mnras/stw137}, 457, 2273

\bibitem[\protect\citeauthoryear{Osborn et~al.,}{Osborn et~al.}{2020}]{osborn2020rapid}
Osborn H.~P.,  et~al., 2020, Astronomy \& Astrophysics, 633, A53

\bibitem[\protect\citeauthoryear{Pollack, Hubickyj, Bodenheimer, Lissauer, Podolak  \& Greenzweig}{Pollack et~al.}{1996}]{pollack1996formation}
Pollack J.~B.,  Hubickyj O.,  Bodenheimer P.,  Lissauer J.~J.,  Podolak M.,   Greenzweig Y.,  1996, Icarus, 124, 62

\bibitem[\protect\citeauthoryear{Rao, Mahabal, Rao  \& Raghavendra}{Rao et~al.}{2021}]{rao2021nigraha}
Rao S.,  Mahabal A.,  Rao N.,   Raghavendra C.,  2021, \mn@doi [Monthly Notices of the Royal Astronomical Society] {10.1093/mnras/stab203}, 502, 2845

\bibitem[\protect\citeauthoryear{Rauer, Aerts, Cabrera  \& {PLATO Team}}{Rauer et~al.}{2016}]{rauer2016plato}
Rauer H.,  Aerts C.,  Cabrera J.,   {PLATO Team} 2016, Astronomische Nachrichten, 337, 961

\bibitem[\protect\citeauthoryear{Ricker et~al.,}{Ricker et~al.}{2015}]{ricker2015transiting}
Ricker G.~R.,  et~al., 2015, Journal of Astronomical Telescopes, Instruments, and Systems, 1, 014003

\bibitem[\protect\citeauthoryear{{Rojas} et~al.,}{{Rojas} et~al.}{2026}]{rojas2026}
{Rojas} F.~I.,  et~al., 2026, \mn@doi [arXiv e-prints] {10.48550/arXiv.2602.13125}, \href {https://ui.adsabs.harvard.edu/abs/2026arXiv260213125R} {p. arXiv:2602.13125}

\bibitem[\protect\citeauthoryear{Salinas, Pichara, Brahm, Pérez-Galarce  \& Mery}{Salinas et~al.}{2023}]{salinas2023}
Salinas H.,  Pichara K.,  Brahm R.,  Pérez-Galarce F.,   Mery D.,  2023, \mn@doi [Monthly Notices of the Royal Astronomical Society] {10.1093/mnras/stad1173}, 522, 3201

\bibitem[\protect\citeauthoryear{Salinas, Brahm, Olmschenk, Barry, Pichara, Ishitani Silva  \& Araujo}{Salinas et~al.}{2025}]{salinas2025}
Salinas H.,  Brahm R.,  Olmschenk G.,  Barry R.~K.,  Pichara K.,  Ishitani Silva S.,   Araujo V.,  2025, \mn@doi [Monthly Notices of the Royal Astronomical Society] {10.1093/mnras/staf347}, 538, 2031

\bibitem[\protect\citeauthoryear{Seager \& Mall{\'e}n-Ornelas}{Seager \& Mall{\'e}n-Ornelas}{2003}]{seager2003unique}
Seager S.,  Mall{\'e}n-Ornelas G.,  2003, \mn@doi [The Astrophysical Journal] {10.1086/346105}, 585, 1038

\bibitem[\protect\citeauthoryear{Snyder, Garcia-Romero, Sell, Povey  \& Khudanpur}{Snyder et~al.}{2018}]{snyder2018x}
Snyder D.,  Garcia-Romero D.,  Sell G.,  Povey D.,   Khudanpur S.,  2018, in 2018 IEEE International Conference on Acoustics, Speech and Signal Processing (ICASSP). pp 5329--5333

\bibitem[\protect\citeauthoryear{Spergel et~al.,}{Spergel et~al.}{2015}]{spergel2015wide}
Spergel D.,  et~al., 2015, \mn@doi [arXiv e-prints] {10.48550/arXiv.1503.03757}, p. arXiv:1503.03757

\bibitem[\protect\citeauthoryear{Srivastava, Hinton, Krizhevsky, Sutskever  \& Salakhutdinov}{Srivastava et~al.}{2014}]{srivastava2014dropout}
Srivastava N.,  Hinton G.,  Krizhevsky A.,  Sutskever I.,   Salakhutdinov R.,  2014, Journal of Machine Learning Research, 15, 1929

\bibitem[\protect\citeauthoryear{Stassun et~al.,}{Stassun et~al.}{2019}]{stassun2019revised}
Stassun K.~G.,  et~al., 2019, \mn@doi [The Astronomical Journal] {10.3847/1538-3881/ab3467}, 158, 138

\bibitem[\protect\citeauthoryear{Tenenbaum et~al.,}{Tenenbaum et~al.}{2012}]{tenenbaum2012detection}
Tenenbaum P.,  et~al., 2012, \mn@doi [The Astrophysical Journal Supplement Series] {10.1088/0067-0049/199/1/24}, 199, 24

\bibitem[\protect\citeauthoryear{Tey et~al.,}{Tey et~al.}{2023}]{tey2023identifying}
Tey E.,  et~al., 2023, \mn@doi [The Astronomical Journal] {10.3847/1538-3881/acad85}, 165, 95

\bibitem[\protect\citeauthoryear{Ulmer-Moll et~al.,}{Ulmer-Moll et~al.}{2022}]{ulmer2022two}
Ulmer-Moll S.,  et~al., 2022, \mn@doi [Astronomy \& Astrophysics] {10.1051/0004-6361/202243583}, 666, A46

\bibitem[\protect\citeauthoryear{Wilson et~al.,}{Wilson et~al.}{2023}]{wilson2023transiting}
Wilson R.~F.,  et~al., 2023, \mn@doi [The Astrophysical Journal Supplement Series] {10.3847/1538-4365/acf3df}, 269, 5

\bibitem[\protect\citeauthoryear{Winn}{Winn}{2010}]{winn2010exoplanet}
Winn J.~N.,  2010, in Seager S.,  ed., , Exoplanets.
University of Arizona Press, Tucson, AZ, pp 55--77 (\mn@eprint {arXiv} {1001.2010})

\bibitem[\protect\citeauthoryear{Wong et~al.,}{Wong et~al.}{2020}]{wong2020systematic}
Wong I.,  et~al., 2020, \mn@doi [The Astronomical Journal] {10.3847/1538-3881/ababad}, 160, 155

\bibitem[\protect\citeauthoryear{Wong et~al.,}{Wong et~al.}{2021}]{wong2021visible}
Wong I.,  et~al., 2021, \mn@doi [The Astronomical Journal] {10.3847/1538-3881/ac0c7d}, 162, 127

\bibitem[\protect\citeauthoryear{Wunderlich \& Sklar}{Wunderlich \& Sklar}{2023}]{wunderlich2023data}
Wunderlich A.,  Sklar J.,  2023, \mn@doi [Machine Learning: Science and Technology] {10.1088/2632-2153/acee44}, 4, 035023

\makeatother
\end{thebibliography}

\onecolumn
\small
\begin{longtable}{lccccc}
\caption{Single-transit candidates.}
\label{tab:single_candidates}\\
\toprule  \hline
TIC ID &  & $T_0$ [BJD] & Duration [h] & Depth [ppm] & $R_p$ [$R_J$] \\
\midrule
\endfirsthead
\multicolumn{6}{c}{{\tablename\ \thetable{} -- continued}} \\
\toprule
TIC ID &  & $T_0$ [BJD] & Duration [h] & Depth [ppm] & $R_p$ [$R_J$] \\
\midrule
\endhead
\midrule \multicolumn{6}{r}{{Continued on next page}} \\
\endfoot
\bottomrule
\endlastfoot
\csvreader[late after line=\\]%
  {data/candidates_single.csv}%
  {tic_id=\ticid, sector=\sector,
   t0_pred=\tzero, t0_pred_error=\tzeroerr,
   duration_pred=\dur, duration_pred_error=\durerr,
   depth_pred=\depth, depth_pred_error=\deptherr,
   final_radius=\radius}%
  {\ticid &  & $\tzero \pm \tzeroerr$ & $\dur \pm \durerr$ & $\depth \pm \deptherr$ & \fmtnum{\radius}}
\end{longtable}

\begin{longtable}{lccccc}
\caption{Multi-transit candidates.}
\label{tab:multi_transit_candidates}\\
\toprule  \hline
TIC ID & Period [d] & $T_0$ [BJD] & Duration [h] & Depth [ppm] & $R_p$ [$R_J$] \\
\midrule
\endfirsthead
\multicolumn{6}{c}{{\tablename\ \thetable{} -- continued}} \\
\toprule
TIC ID & Period [d] & $T_0$ [BJD] & Duration [h] & Depth [ppm] & $R_p$ [$R_J$] \\
\midrule
\endhead
\midrule \multicolumn{6}{r}{{Continued on next page}} \\
\endfoot
\bottomrule
\endlastfoot
\csvreader[late after line=\\]%
  {data/candidates_multi_transit.csv}%
  {tic_id=\ticid, period=\period,
   t0_pred=\tzero, t0_pred_error=\tzeroerr,
   duration_pred=\dur, duration_pred_error=\durerr,
   depth_pred=\depth, depth_pred_error=\deptherr,
   final_radius=\radius}%
  {\ticid & \fmtnum{\period} & $\tzero \pm \tzeroerr$ & $\dur \pm \durerr$ & $\depth \pm \deptherr$ & \fmtnum{\radius}}
\end{longtable}

\begin{longtable}{lccccc}
\caption{Multi-planet candidates.}
\label{tab:multi_planet_candidates}\\
\toprule  \hline
TIC ID & & $T_0$ [BJD] & Duration [h] & Depth [ppm] & $R_p$ [$R_J$] \\
\midrule
\endfirsthead
\multicolumn{6}{c}{{\tablename\ \thetable{} -- continued}} \\
\toprule
TIC ID & & $T_0$ [BJD] & Duration [h] & Depth [ppm] & $R_p$ [$R_J$] \\
\midrule
\endhead
\midrule \multicolumn{6}{r}{{Continued on next page}} \\
\endfoot
\bottomrule
\endlastfoot

\csvreader[late after line=\\]%
  {data/candidates_multi_planet.csv}%
  {tic_id=\ticid, nro_planet=\nrop,
   t0_pred=\tzero, t0_pred_error=\tzeroerr, duration_pred=\dur, duration_pred_error=\durerr, depth_pred=\depth,depth_pred_error=\deptherr,
   final_radius=\radius}%
  {\ticid & \nrop & $\tzero \pm \tzeroerr$ & $\dur \pm \durerr$ & $\depth \pm \deptherr$ & \fmtnum{\radius}}
\end{longtable}
\twocolumn


\clearpage

\appendix
\onecolumn  

\raggedbottom

\section{Appendix}
\label{ap:appendix}
This appendix presents supplementary figures and tables of the results discussed in the main text, including the training dataset used to develop the CNN and the full per-level sensitivity results referenced in Section~\ref{sec:results}.

\begin{center}
  \includegraphics[width=\linewidth]{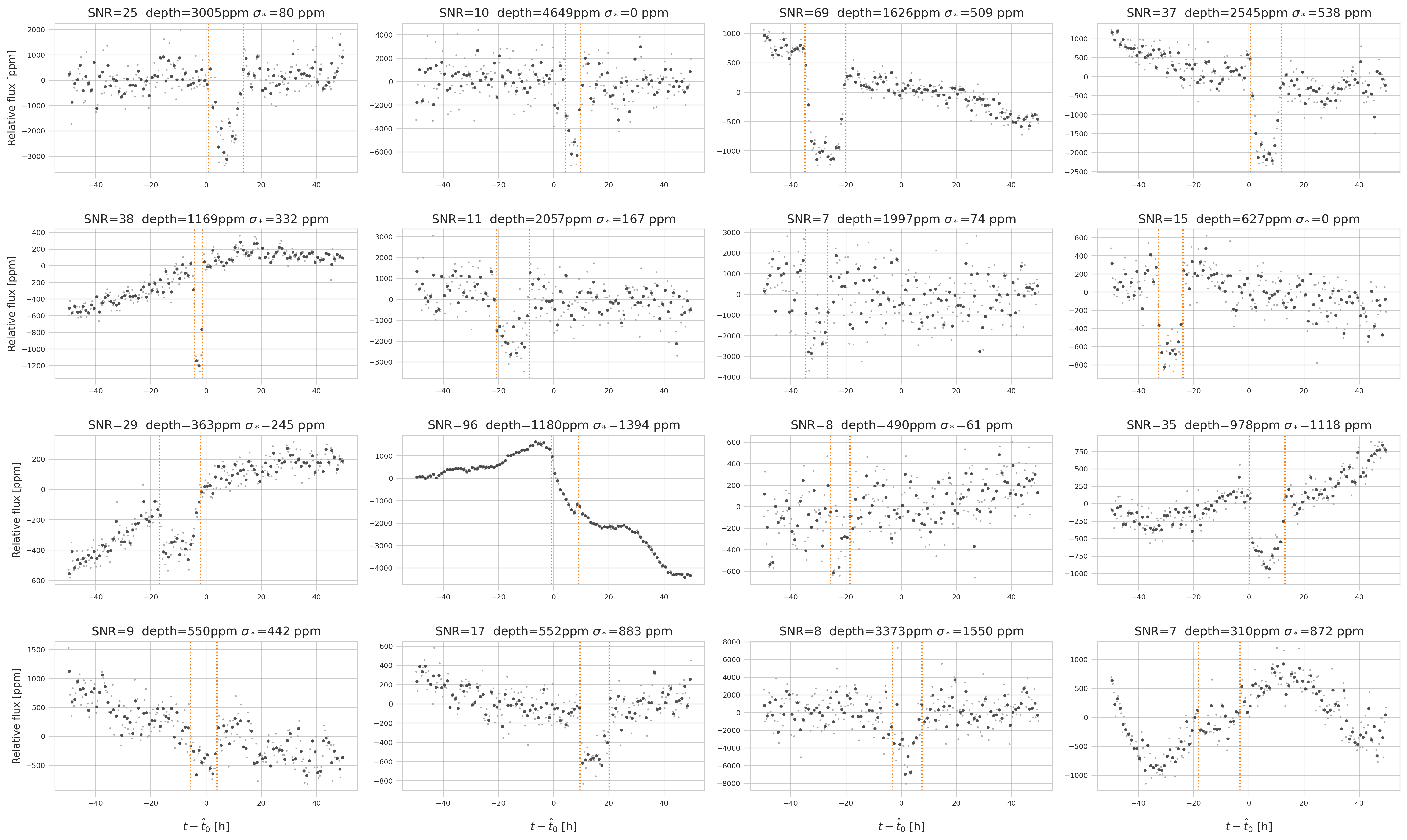}
  \captionof{figure}{Sample of synthetic LCs from the training set, showing the range of SNR, transit depth, and stellar variability. Dotted vertical lines mark the true transit edges.}
  \label{fig:dataset_sim}
\end{center}


\newcommand{\sensrows}[2]{%
  \csvreader[
    late after line=\\,
    filter strcmp={\test}{#1}
  ]{data/sensitivity_results.csv}{
    show_test=\showtest,
    test=\test,
    level=\level,
    level_unit=\levelunit,
    sigma_w_median_ppm=\sigmaw,
    snr_median=\snr,
    mae_depth_ppm=\maedepth,
    mae_dur_h=\maedur,
    dt_fail_pct=\dtfail
  }{%
    \ifnum\showtest=1\relax #2\fi
    & \level & \levelunit & \sigmaw & \snr & \maedepth & \maedur & \dtfail
  }
}

\begin{longtable}{llcccccc}
\caption{Sensitivity tests. Each set contains 2000 synthetic LCs. $\sigma_w$ median denotes the median white noise across the set and $\Delta t$ fail is the fraction with epoch error $>5$ h.}
\label{tab:sensitivity}\\
\hline\hline
Parameter & Level & Unit & $\sigma_w$ median & SNR median & MAE depth & MAE duration & $\Delta t$ fail \\
 & & & [ppm] & & [ppm] & [h] & [\%] \\
\hline
\endfirsthead
\multicolumn{8}{c}{{\tablename\ \thetable{} -- continued}} \\
\hline\hline
Parameter & Level & Unit & $\sigma_w$ median & SNR median & MAE depth & MAE duration & $\Delta t$ fail \\
 & & & [ppm] & & [ppm] & [h] & [\%] \\
\hline
\endhead
\hline \multicolumn{8}{r}{{Continued on next page}} \\
\endfoot
\hline\hline
\endlastfoot
\sensrows{White noise}{White noise $\sigma_w$}
\hline
\sensrows{Red noise}{Red noise $\sigma_r$}
\hline
\sensrows{Stellar var.}{Stellar var. $\sigma_\star$}
\hline
\sensrows{Epoch offset}{Epoch offset $\Delta t_0$}
\hline
\sensrows{Impact param.}{Impact parameter $b_{\rm max}$}
\end{longtable}

\begin{center}
  \includegraphics[width=0.9\linewidth]{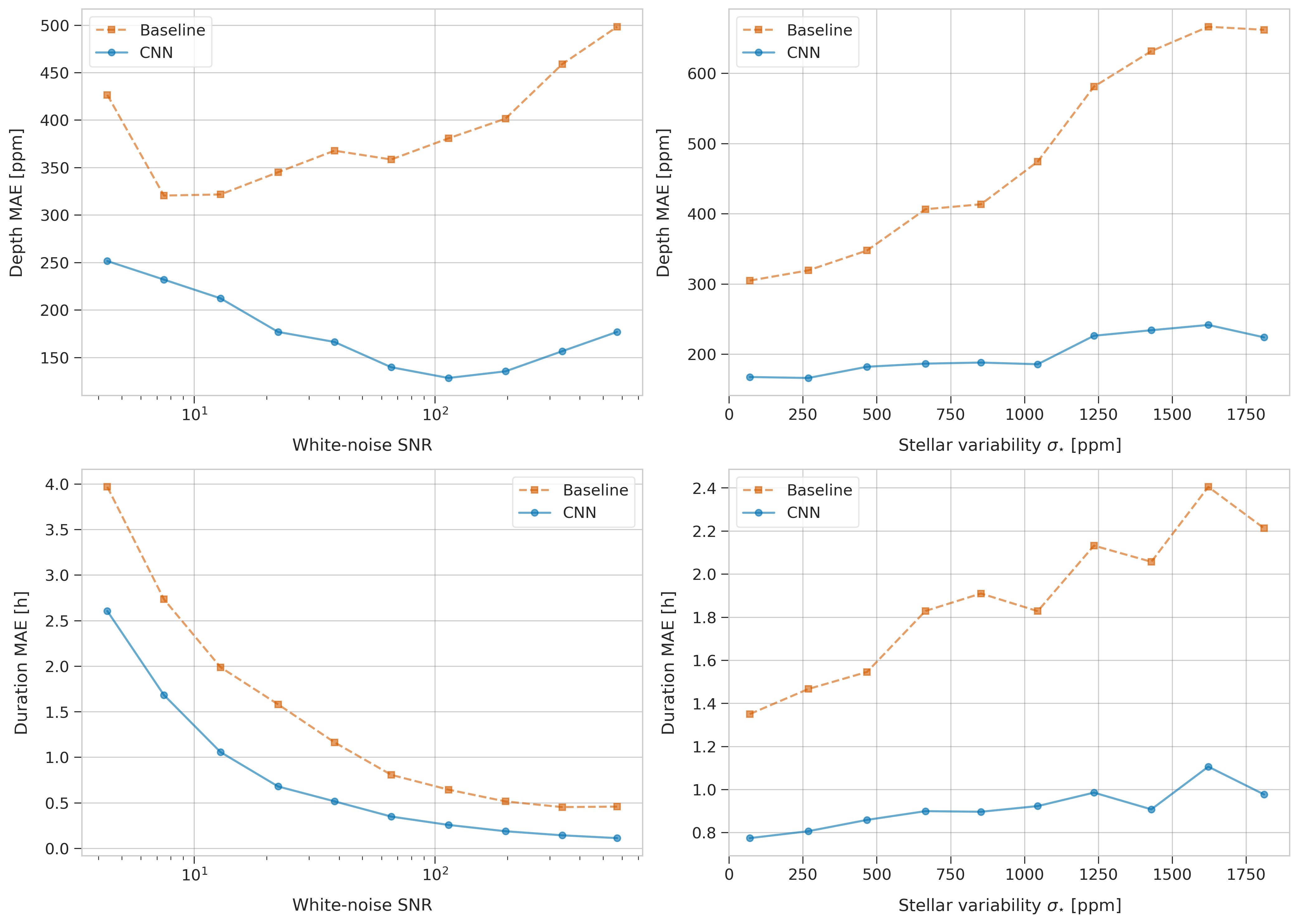}
  \captionof{figure}{Comparison of CNN and Trapezoidal-LS baseline performance on the simulated data as a function of white-noise SNR (left column) and stellar variability amplitude $\sigma_\star$ (right column), for depth MAE (top) and duration MAE (bottom).}
  \label{fig:mae_depth_dur_cnn_vs_ls}
\end{center}

\begin{center}
  \includegraphics[width=1.\linewidth]{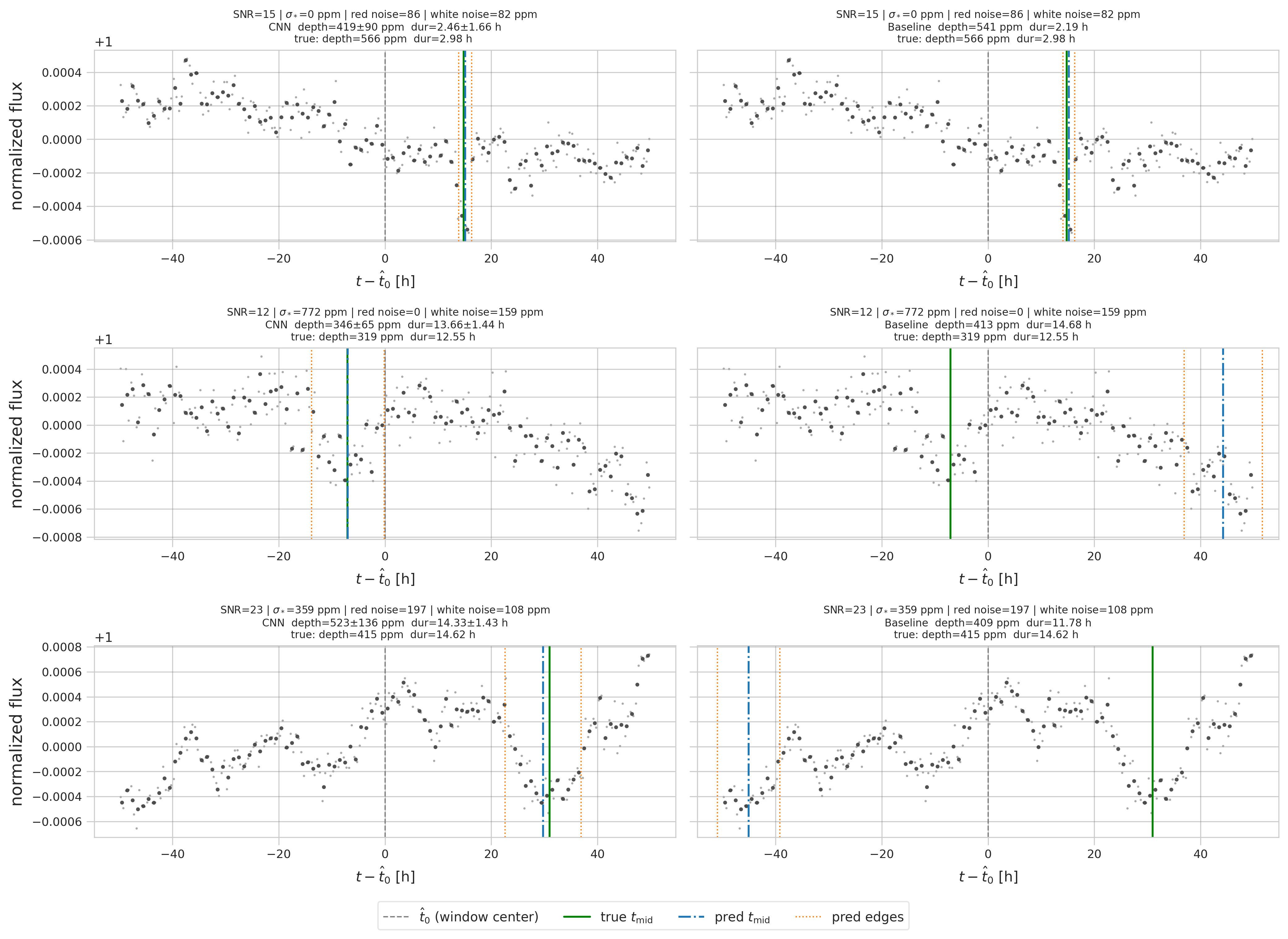}
  \captionof{figure}{Comparison of CNN and Trapezoidal-LS baseline predictions for three synthetic LCs. 
  Left column: CNN predictions with $1\sigma$ uncertainties. Right column: Trapezoidal-LS baseline predictions. In each panel, the dashed vertical line marks the window centre $\hat{t}_0$, the solid vertical line marks the true mid-transit time, the dot-dashed line marks the predicted mid-transit time, and the dotted lines 
  mark the predicted transit edges. Predicted depth and duration, together with the true values and noise properties, are annotated above each panel.}
  \label{fig:prediction_sim_data}
\end{center}




\flushbottom
\bsp	
\label{lastpage}
\end{document}